\documentclass[a4paper,fleqn,usenatbib]{mnras}

\usepackage[T1]{fontenc}
\usepackage{ae,aecompl}

\usepackage{graphicx}	% Including figure files
\usepackage{amsmath}	% Advanced maths commands
\usepackage{amssymb}	% Extra maths symbols
\usepackage{multirow}
\usepackage{booktabs}
\usepackage{rotating}
\usepackage{xcolor}

\newcommand{\nicer}{\textit{NICER}}
\newcommand{\nustar}{\textit{NuSTAR}}

\newcommand{\swift}{{\it Swift}}

\newcommand{\maxi}{MAXI~J1820+070}

\title[Soft-state flaring of \maxi\ with a long soft lag]{\maxi\ with \nustar\ II. Flaring during the hard to soft state transition with a long soft lag}

\author[D. J. K. Buisson et al.]{D. J. K. Buisson$^{1,2}$,\thanks{Email: d.j.k.buisson@soton.ac.uk}
	A. C. Fabian$^{2}$,
	P. Gandhi$^{1}$,
	E. Kara$^{3}$,
	M. L. Parker$^{4}$,
\newauthor	A. W. Shaw$^{5}$,
    J. A. Tomsick$^{6}$,
    D. J. Walton$^{2}$ and
    J. Wang$^{3}$
\\
  $^{1}$Department of Physics and Astronomy, University of Southampton, Highfield, Southampton, SO17 1BJ\\
  $^{2}$Institute of Astronomy, Madingley Road, Cambridge, CB3 0HA\\
  $^{3}$MIT Kavli Institute for Astrophysics and Space Research, Massachusetts Institute of Technology, Cambridge, MA 02139, USA\\
  $^{4}$European Space Astronomy Centre (ESA/ESAC), E-28691 Villanueva de la Ca\~nada, Madrid, Spain\\
  $^{5}$Department of Physics, University of Nevada, Reno, NV 89557, USA\\
  $^{6}$Space Sciences Laboratory, 7 Gauss Way, University of California, Berkeley, CA 94720-7450, USA\\
}

\date{Accepted XXX. Received YYY; in original form ZZZ}

\pubyear{2020}

\begin{document}
\label{firstpage}
\pagerange{\pageref{firstpage}--\pageref{lastpage}}
\maketitle

\begin{abstract}
We continue the analysis of \nustar\ data from the recent discovery outburst of \maxi\ (optical counterpart ASASSN-18ey), focussing on an observation including unusual flaring behaviour during the hard to soft state transition, which is a short phase of outbursts and so comparatively rarely observed.
Two plateaus in flux are separated by a variable interval lasting $\sim10$\,ks, which shows dipping then flaring stages.
The variability is strongest (with fractional variability up to $F_{\rm Var}\sim10\%$) at high energies and reduces as the contribution from disc emission becomes stronger.
Flux-resolved spectra show that the variability is primarily due to the power law flux changing.
We also find a long soft lag of the thermal behind the power law emission, which is $20_{-1.2}^{+1.6}$\,s during the flaring phase.
The lag during the dipping stage has a different lag-energy spectrum, which may be due to a wave passing outwards through the disc.
Time resolved spectral fitting suggests that the lag during the flaring stage may be due to the disc re-filling after being disrupted to produce the power law flare, perhaps related to the system settling after the jet ejection which occurred around 1 day before.
The timescales of these phenomena imply a low viscosity parameter, $\alpha\sim10^{-3}$, for the inner region of the disc.
\end{abstract}

\section{Introduction}

X-ray binaries (XRBs) are systems in which a compact object (neutron star or black hole) accretes from its companion star, releasing energy across the electromagnetic spectrum, including substantial power in X-rays.
While fundamentally powered simply by gravity, the emission in XRBs displays a rich phenomenology.

There are two principal components in the X-ray emission of XRBs: thermal emission from the disc around the compact object \citep{novikov73,shakura73} and Comptonised emission from a hotter, optically thin region of plasma known as the corona \citep{thorne75,sunyaev79,haardt91}.
For most of the active period of XRBs, one of these components produces a large majority of the emitted power, and transitions between periods when the disc dominates and periods when the corona dominates are comparatively rapid \citep[e.g. review by][]{remillard06}.

The reason for these state changes is not yet well understood.
One common model is that the inner region of the accretion disc changes from a hot, optically thin, highly ionised flow in the hard state and to optically thick and thermally emitting in the soft state \citep{esin97,done07,gilfanov10}. The hot inner flow then emits through Comptonisation.
However, the inner radius of the disc is often found to be too small for this to be the case \citep{park04,reis13,parker15,buisson19}.
Instead, the inner region of the disc may only emit a small fraction of its energy thermally \citep{reis10}, while the remainder is extracted by other, probably magnetic, processes, and powers a corona elsewhere, such as the base of the jet \citep[e.g.][]{markoff05,fabian12}.

Since X-Ray Binaries (XRBs) cannot currently be resolved spatially, other means must be used to determine the physical structure: variability is an important tool in determining both the processes (through the amount of variability at different energies) and  geometry (through echoes between different components) involved in their emission.
The variability observed in XRBs has been classified into a wide variety of phenomena, which are seen at different points in the outburst.
The majority of variability is associated with the Comptonised emission so variability is principally observed at times when and energies where this component dominates the X-ray spectrum.

Typically, much of the variability is in the form of broadband noise: this shows relatively little structure in the light curve and occurs over a wide range of Fourier frequencies, \citep[e.g.][]{belloni05}.
While there is little structure in the amount of variability at different time scales, there are particular lag properties between different energies.
On short time scales ($\gtrsim1$\,Hz), there are lags at soft energies and around the iron-K line due to the extra light travel time of emission which is reflected from the disc \citep{uttley14,demarco16,kara19}.
On longer time scales, emission at harder energies lags that at softer \citep{miyamoto89}; this is thought to be due to the propagation of fluctuations through the disc \citep{uttley11}.

Often, variability is particularly strong on particular timescales; these are Quasi Periodic Oscillations (QPOs; e.g. \citealt{vanderklis06}) 
and are further divided into subtypes based on their frequency, coherence and the strength of different harmonics \citep{wijnands99,homan01,remillard02}.

Occasionally, variability may be much more structured. For example, the Rapid Burster \citep[MXB~1730--335, e.g.][]{hoffman78} and Bursting Pulsar  (GRO~J1744--28, \citealt{fishman95,kouveliotou96}) show strong, brief increases in flux referred to as Type~II X-ray bursts.
Other sources occasionally show fast, repetitive transitions between two distinct emission regimes \citep{miyamoto91,bogensberger20} referred to as `flip-flops'.
The most extreme case of structured variability is seen in GRS~1915+105, where 14 distinct variability modes have been classified \citep{belloni00,kleinwolt02,hannikainen05}; such complex variability is observed in no other source, apart perhaps from IGR~J17091--3624 \citep{altamirano11,court17}.

\subsection{\maxi}

\begin{figure*}
\centering
\includegraphics[ trim={0.cm 0 0 0},width=\columnwidth]{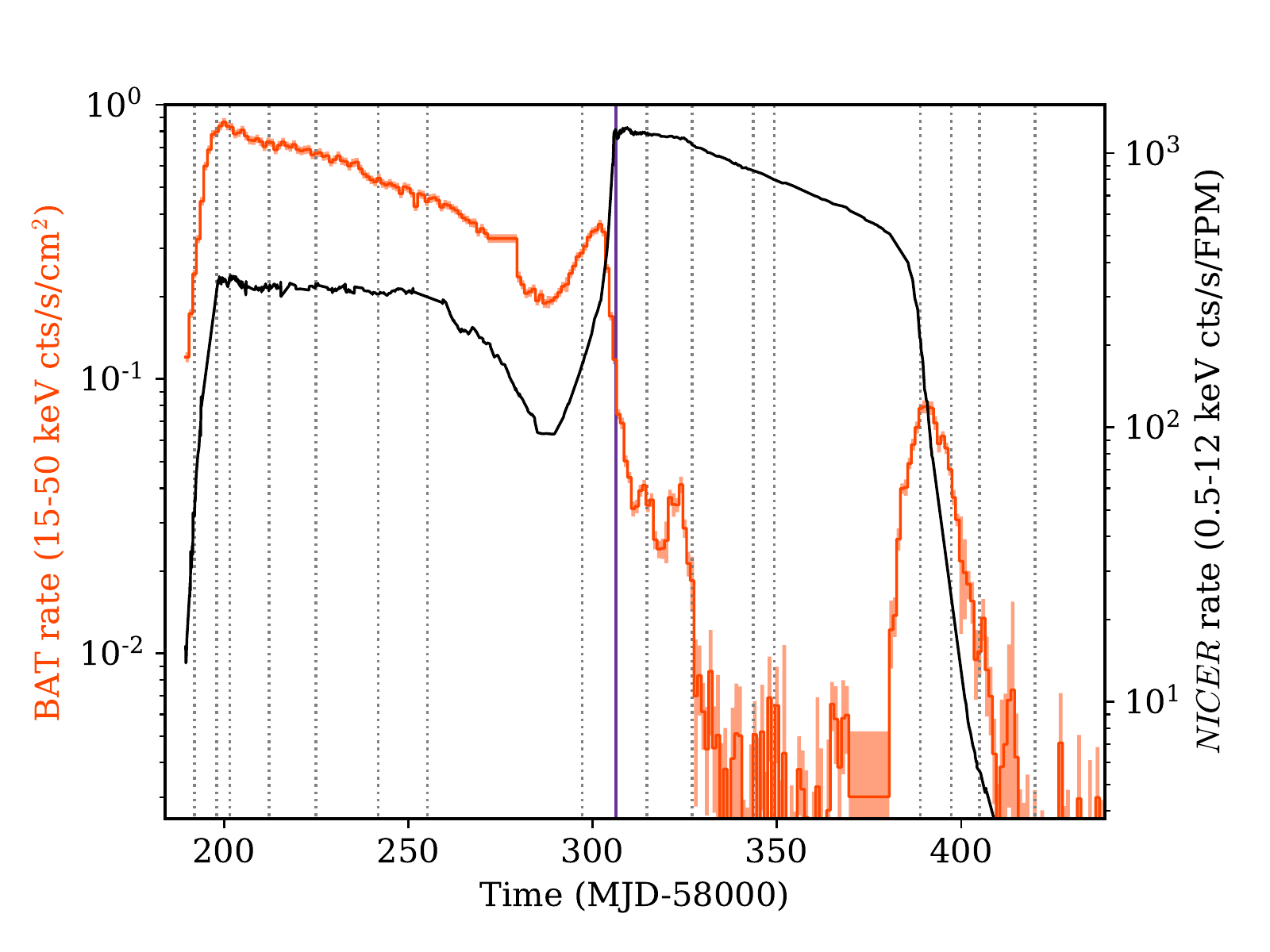}
\includegraphics[ trim={0.cm 0 0 0},width=\columnwidth]{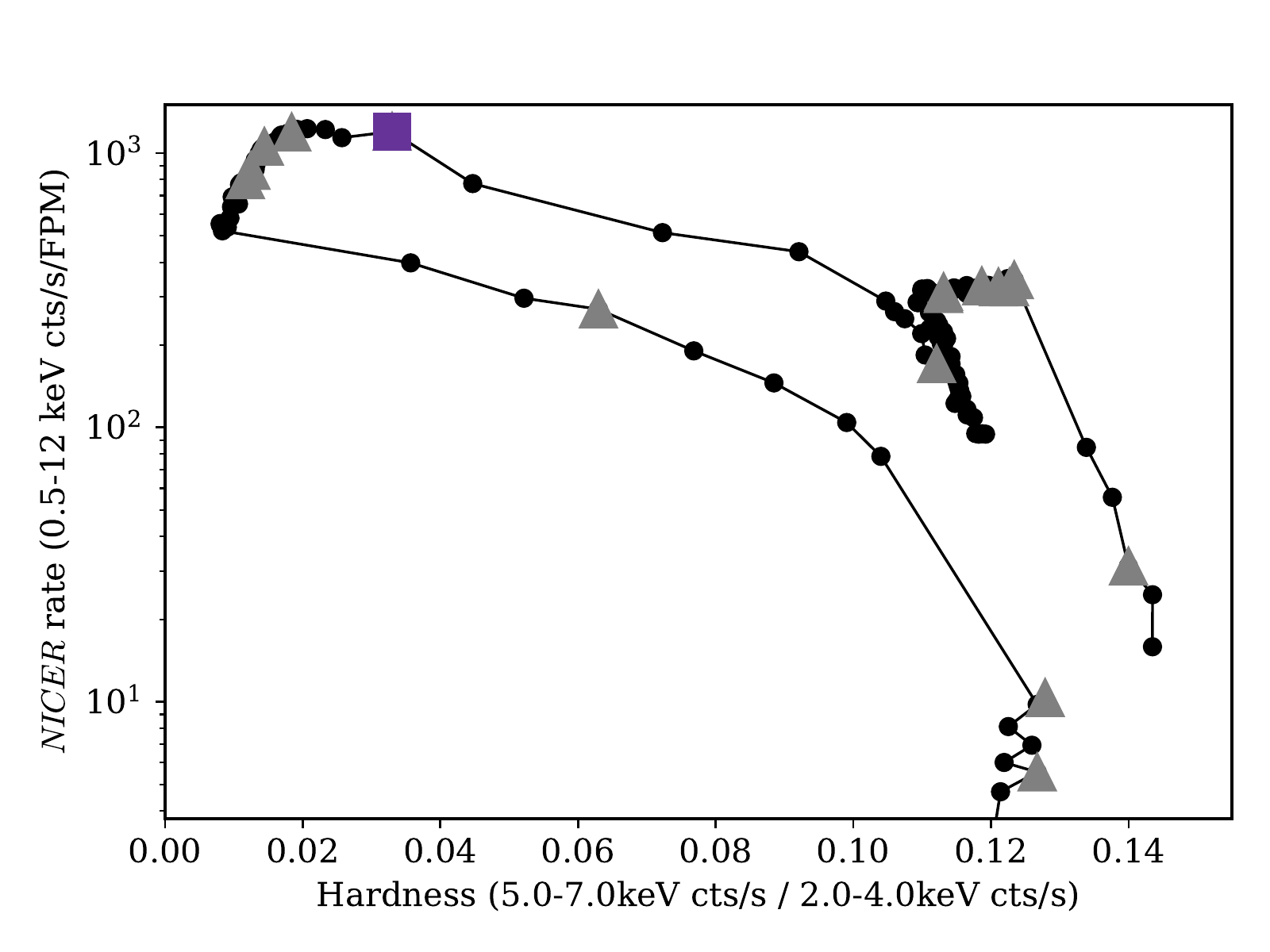}
\includegraphics[ trim={0.cm 0 0 0},width=\columnwidth]{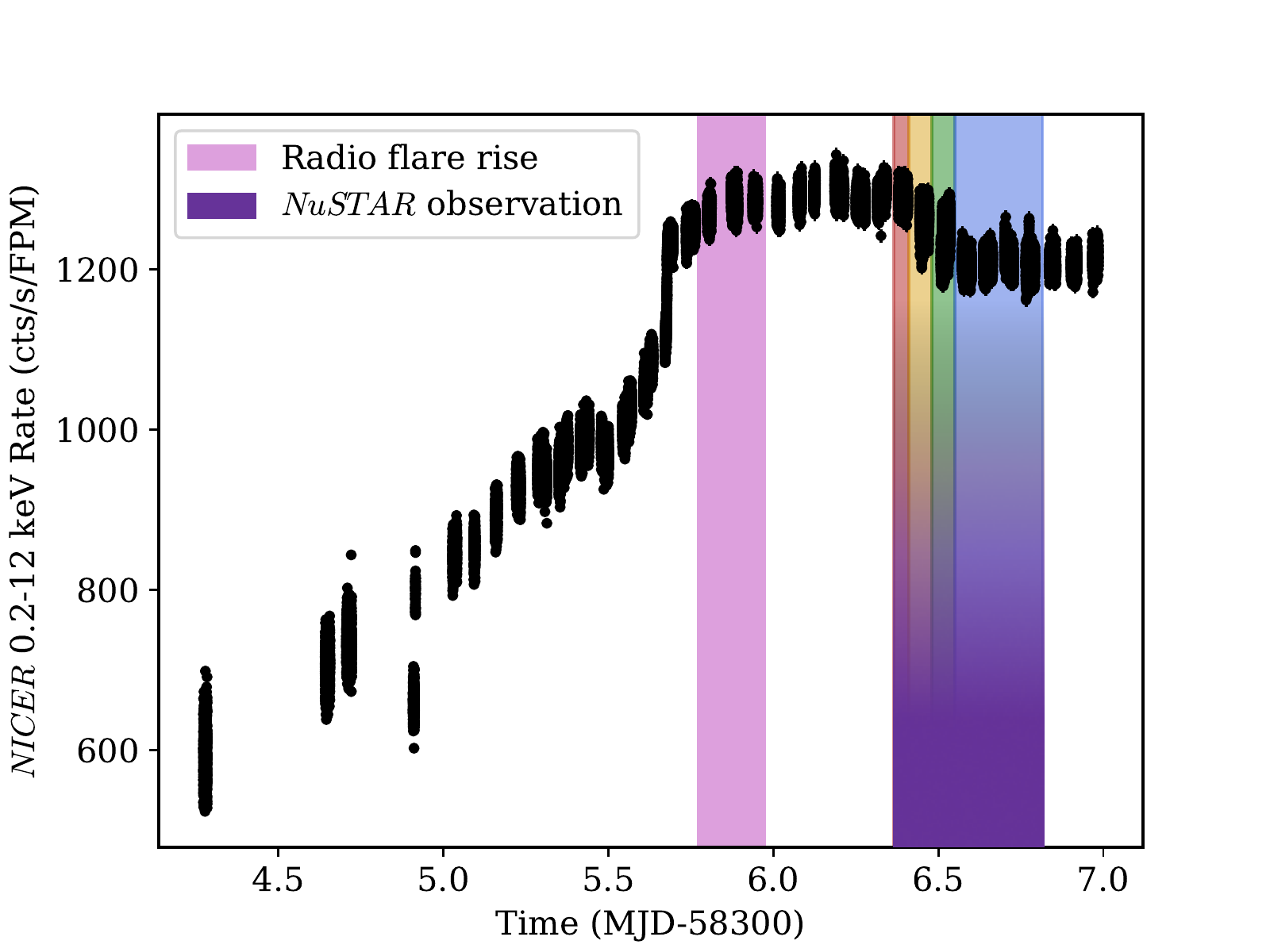}
\includegraphics[width=\columnwidth]{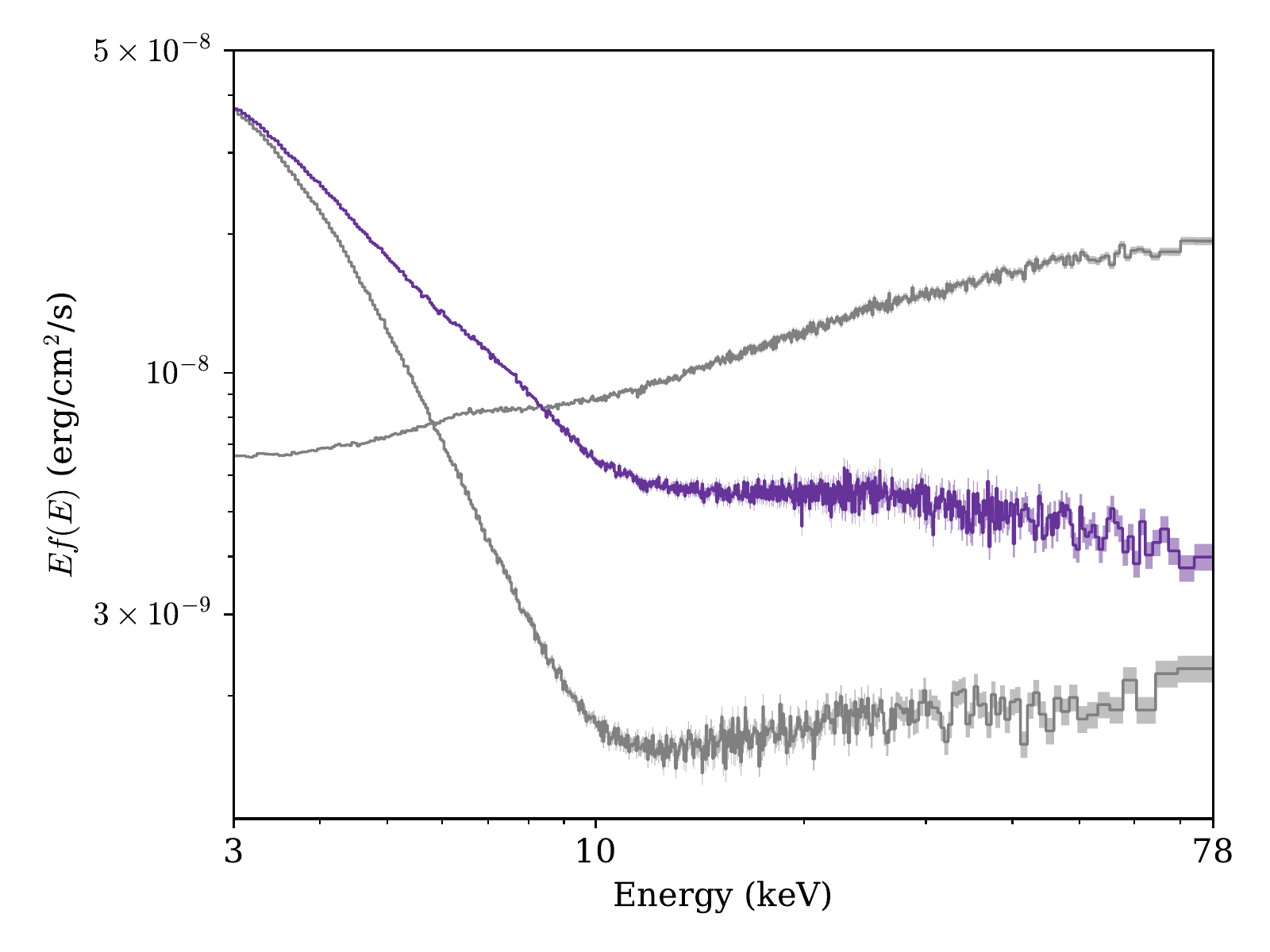}
\caption{{\it Top left:} Light curve of the outburst of \maxi\ from \swift-BAT (orange) and \nicer\ (black) with times of \nustar\ observations shown as vertical bars. The observation analysed here is shown in solid purple.
{\it Bottom left:} Light curve of \nicer\ data (black, binned to 1\,s) covering the interval around the state transition. The rise of the radio flare (pink) is slightly ($\approx0.1$\,day) after the strongest change in soft flux and $\approx0.4$\,days before the \nustar\ observation (purple and colours matching Figure~\ref{fig:lc}).
{\it Top right:} Hardness-Intensity diagram of \nicer\ data, averaged per day, with times of \nustar\ observations indicated by grey triangles and the \nustar\ observation analysed here indicated by the purple square.
{\it Bottom right:} Mean spectra of the \nustar\ observation analysed here (purple) and those immediately before (black, hard) and after (black, soft) unfolded to a constant model (i.e. corrected for the instrument's effective area by comparison to a $\Gamma=2$ power law, but without instrumental broadening deconvolved). FPMA and B have been combined for display purposes.}
\label{fig_lc_comp}
\label{fig_unfold}
\end{figure*}

\maxi\ is a black-hole X-ray binary discovered in 2018 in the optical by the All-Sky Automated Search for SuperNovae as ASASSN-18ey \citep{shappee14,tucker18} and X-rays by \textit{MAXI} \citep{kawamuro18,denisenko18}.
It has provided an excellent target for study due to its high brightness and low absorption column.
It was confirmed to host a black hole by \citet{torres19}. Further measurements \citep{torres20} refine the mass estimate to $5.7<M_{\rm BH}/M_{\odot}<9.5$, depending on the assumed binary inclination.
The most precise distance measured to date is $2.96\pm0.33$\,kpc from radio parallax \citep{atri20}; this is consistent with the optical parallax from \textit{GAIA} \citep[$3.5_{-1.0}^{+2.2}$\,kpc,][]{bailerjones18,gandhi19}.

Its outburst began, as is typical, in the hard state but declined to around one quarter of the peak flux before re-brightening and undergoing a transition to the soft state \citep{homan18_transition1a,homan18_transition1b}.
The \nustar\ data from the initial hard state are analysed in \citet{buisson19}, showing that the evolution is governed by changes in coronal rather than disc geometry, agreeing with the trend in reverberation lags seen in \nicer\ data \citep{kara19}.
Several phenomena during the transition have already been reported, including a bipolar radio plasma ejection \citep{bright20} and a type B QPO \citep{homan20}. The jet ejecta were also observed in X-rays \citep{espinasse20}.
\nustar\ observations during the soft state are analysed in \citet{fabian20}, showing excess emission at energies above the disc emission, interpreted as emission from the plunge region inside the Innermost Stable Circular Orbit (ISCO).
After around 80\,days in the soft state, it returned to the hard state \citep{homan18_transition2} while fading towards quiescence. The long-term light curve and hardness-intensity diagram of this period is shown in Figure~\ref{fig_lc_comp}.
Since the main outburst, there have been several smaller outbursts \citep{vozza19,xu19atel,sasaki20}, which remained in the hard state.

Here, we describe the spectra and variability of the X-ray emission during a \nustar\ observation during the hard to soft state transition. For context, we show the \nicer\ light curve around the transition and the mean spectra of this observation along with the previous and following \nustar\ observations in Figure~\ref{fig_unfold}.

\section{Observations and Data Reduction}
\label{section_datareduction}

We analyse data from the first \nustar\ \citep{harrison13} observation of \maxi\ in the transition to the soft state, OBSID 90401309023; this is epoch 9 in the notation of \citet{buisson19}, which analysed epochs 1-8, and Nu23 in the notation of \citet{fabian20}.
The data were reduced as in \citet{buisson19}, the only non-standard feature of this being the use of the status expression \texttt{"STATUS==b0000xxx00xxxx000"} to avoid photons from the bright source being spuriously flagged as \textsc{`test'}.
We note that the background for this observation is negligible and the illumination of the whole of the detector is source-dominated.
The total length of the observation is 38.1\,ks, of which 21.8/21.5\,ks is on-source time, which has an average live fraction of 0.33 (due to the significant effect of dead time at these high count rates) giving effective exposures of 7.1/7.4\,ks for FPMA/B.

We also consider simultaneous \nicer\ data, from OBSID 1200120198. We use the clean events from the standard \nicer\ filtering, apart from the filter for the minimum number of active detectors, and extract spectra and lightcurves using \textsc{xselect}. We note that due to the extreme brightness of \maxi\ during this observation, most of the detectors were switched off, reducing the effective area significantly.

We also show data from the \textit{Neil Gehrels Swift Observatory} Burst Alert Telescope (\swift-BAT) transient monitor \citep{gehrels04,krimm13} for illustrative purposes; we use the online daily light curve\footnote{https://swift.gsfc.nasa.gov/results/transients/weak/ MAXIJ1820p070/}.

\section{Results}
\label{sec:results}

\subsection{Variability in the light curve}
\label{sec:lc}

\begin{figure*}
\includegraphics[width=\columnwidth]{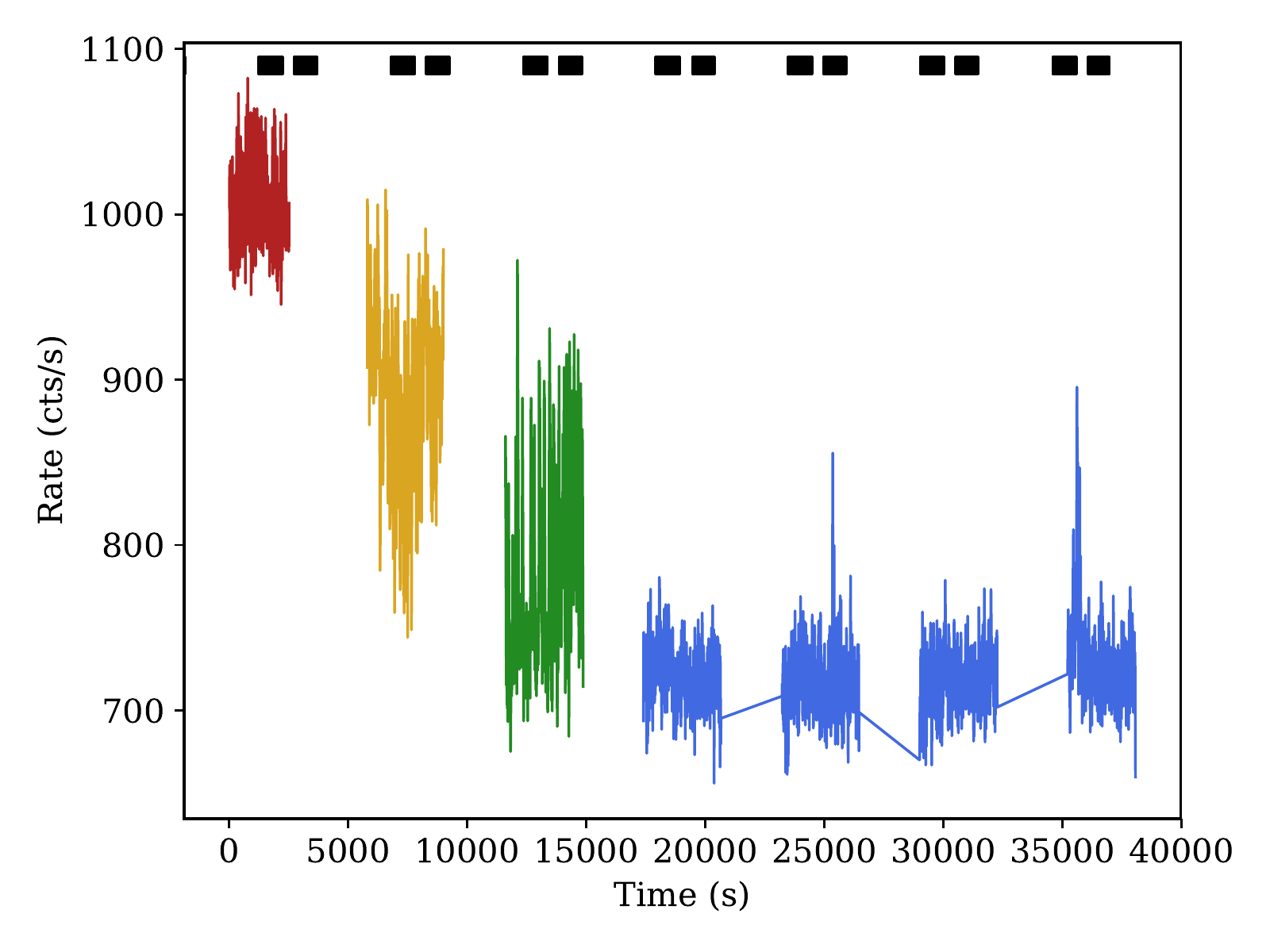}
\includegraphics[width=\columnwidth]{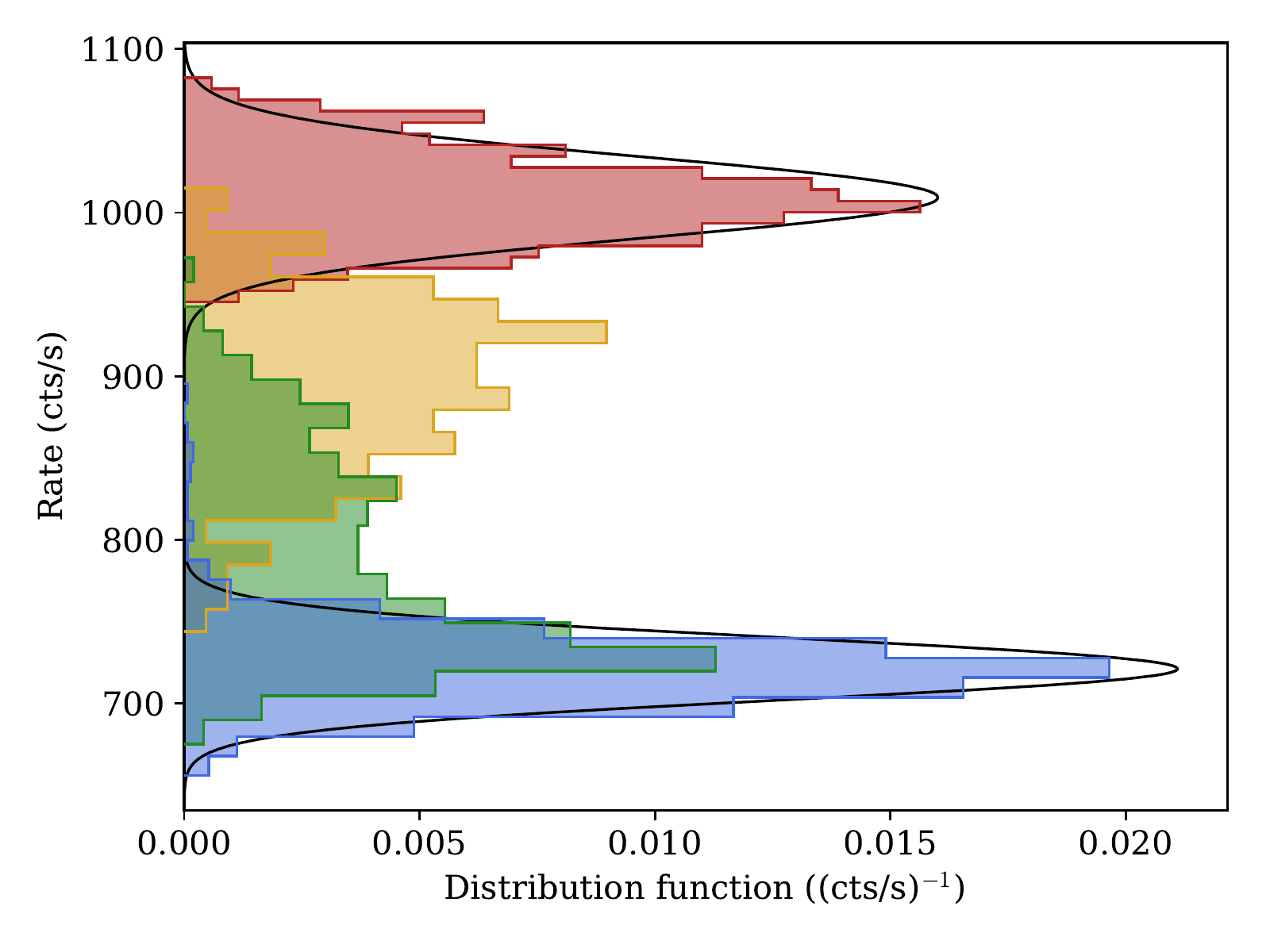}
\includegraphics[width=\columnwidth]{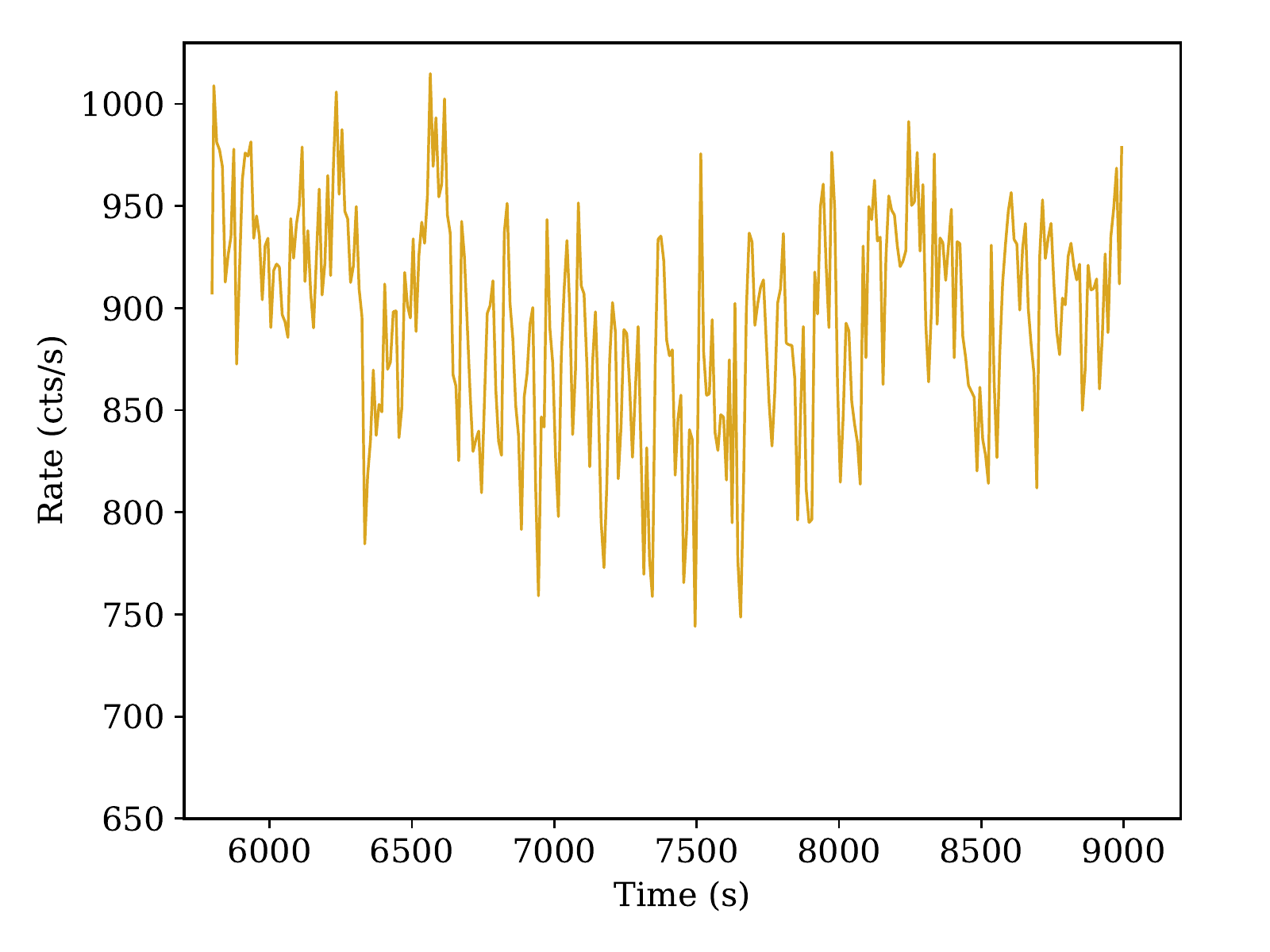}
\includegraphics[width=\columnwidth]{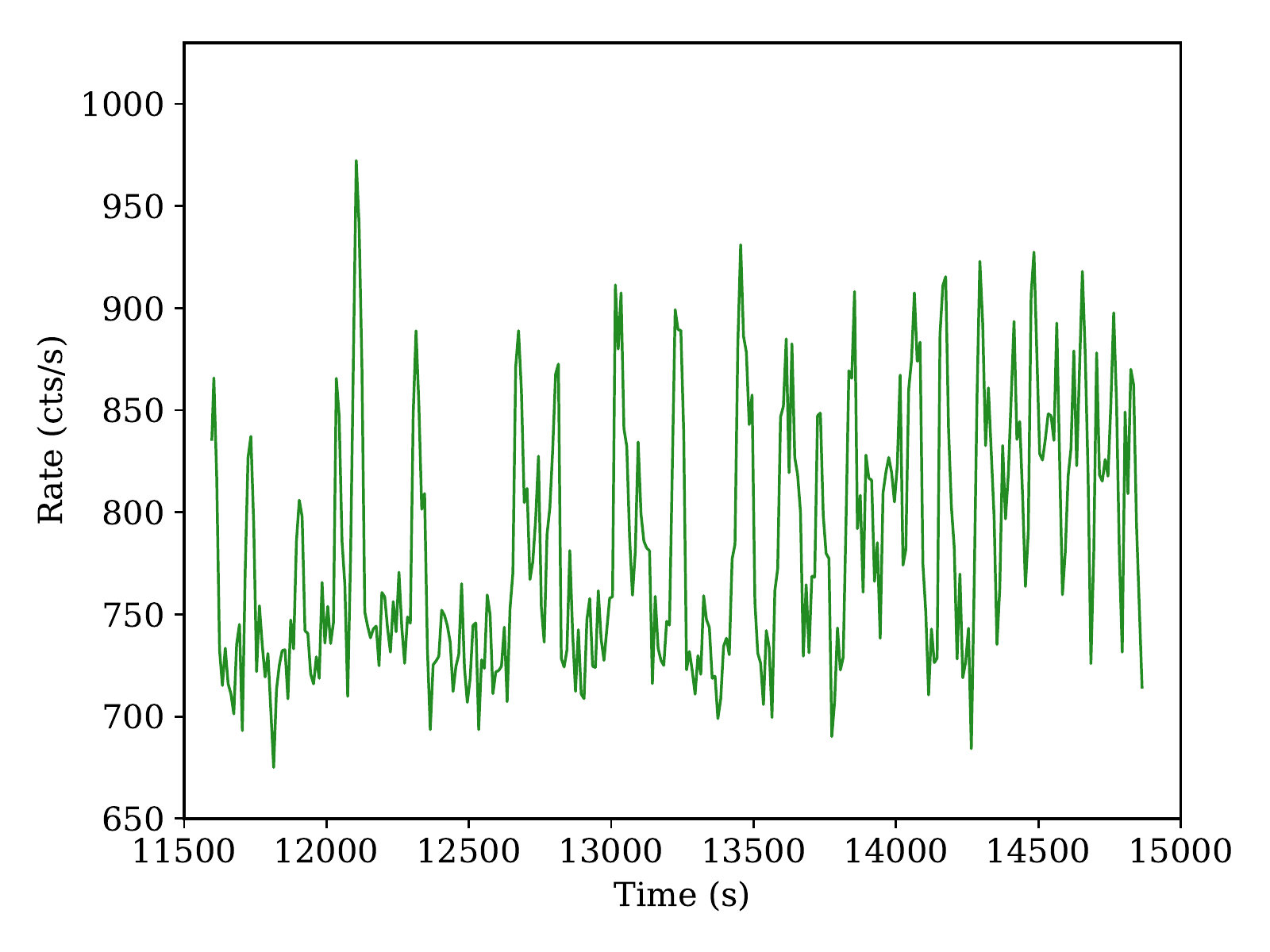}
\caption{\textit{Top left}: $3-78$\,keV \nustar\ light curve, binned to 10\,s. Between two roughly steady sections is a period of significant variability. The black bars at the top of the plot mark the times of simultaneous \nicer\ coverage.
\textit{Top right}: Flux distributions for each light curve section (colours match left panel). The steady sections (red/blue) have width similar to that expected from measurement noise on a constant intrinsic flux (black curves); the first (yellow) variable section has a low flux (dipping) tail; and the second (green) variable section has a peak matching the following steady section with a tail  due to excursions to higher flux.
\textit{Bottom panels}: Zooms to each variable \nustar\ orbit to show variability more clearly.}
\label{fig:lc}
\end{figure*}

We first show the light curve of \maxi\ in the full $3-78$\,keV \nustar\ band (Figure~\ref{fig:lc}). We note that due to the shape of the spectrum and the \nustar\ response, this is dominated by the low energy flux.
This shows a stable flux with little variability in the first orbit and last four orbits (apart from two isolated flares), with the first at considerably ($\sim30$\%) higher flux. The width in the flux distributions of these sections is similar to that expected from measurement noise. Calculating power spectra using the standard formulae \citep[e.g.][]{priestley81} confirms that Poisson measurement noise is dominant, with the only feature being a $\sim9$\,Hz QPO in the final plateau. The amplitude of this QPO is low, requiring the \nicer\ data to detect it. We discuss this feature further in Section~\ref{sec:ff}.
The transition between these two sections shows strong variability, the nature of which appears to differ between the two orbits.
The first variable orbit (yellow) is the brighter of the two and shows little structure to its variability. The flux distribution (Figure~\ref{fig:lc}) shows a longer low than high-flux tail.
Contrastingly, the second variable orbit (green) is highly structured, consisting of flares from a baseline flux similar to that in the following steady period. The flares increase the rate by a factor of $\sim20-25$\% and are spaced by $\sim200$\,s at the start of the orbit, increasing in frequency towards the end of the orbit.
The properties of the two variable orbits are the focus of the remainder of this work; we treat them separately and refer to them as the dipping (yellow) and flaring (green) stages based on the shape of the light curve.

\subsection{Spectral analysis -- full observation}
\label{sec:spec}

\begin{figure}
\centering
\includegraphics[width=\columnwidth]{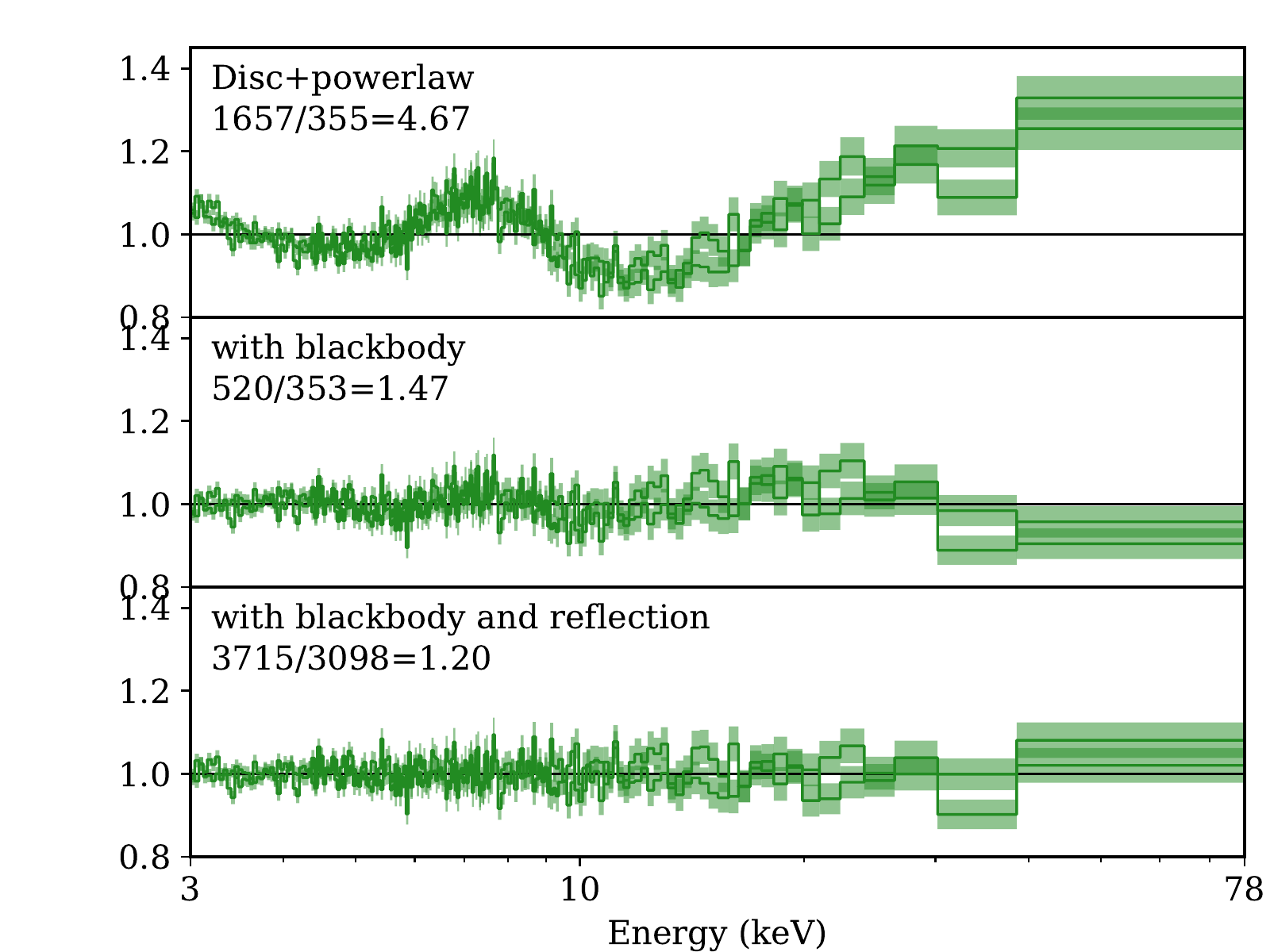}
\caption{Residuals to the models described in the text. Each panel includes a description of the model and the value of $\chi^2/{\rm degrees\ of\ freedom}$. For clarity, only the low flux state of the flaring section is shown. Note that the reflection model is a joint fit across all datasets, hence the significantly greater degrees of freedom.}
\label{fig:res}
\end{figure}

\begin{figure}
\centering
\includegraphics[width=\columnwidth]{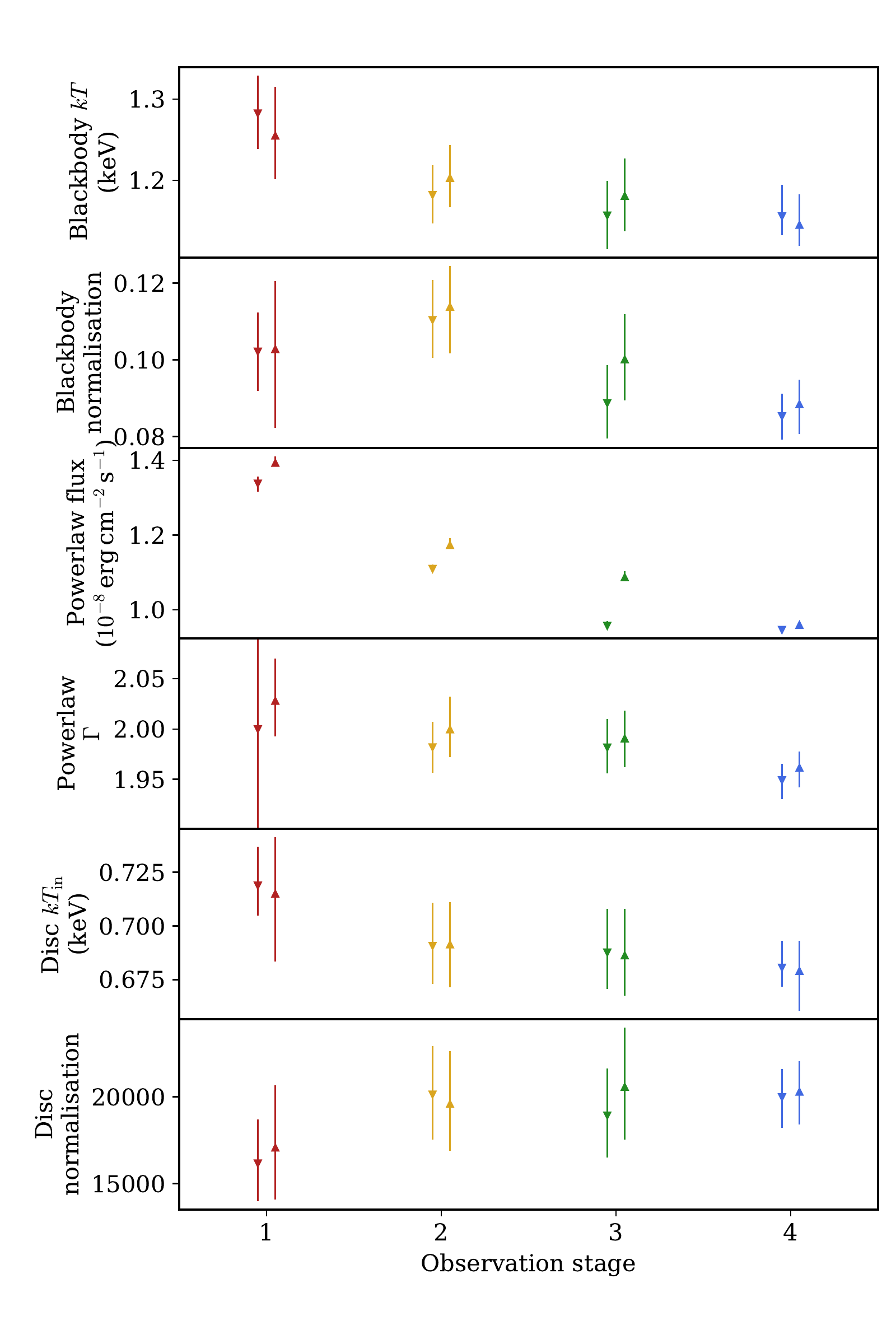}
\caption{Values of fit parameters to flux resolved stages of observation. Colours match those of Figure~\ref{fig:lc}; upwards/downwards pointing triangles mark high/low flux spectra respectively. The largest difference within a stage is due to the power law flux. Across the observation, the thermal components become cooler and the power law becomes fainter and harder.}
\label{fig:pars}
\end{figure}

\begin{table}
\caption{Parameters of reflection component of fit to full \nustar\ observation.}
\label{tab:spec}
\centering
\begin{tabular}{llc}
\hline
\multicolumn{2}{c}{Parameter} & Value\\
\hline
Coronal height & $h/r_{\rm g}$ & $>78$ \\
Spin & $a$ & Unconstrained \\
Inclination & $\theta/^\circ$ & $65\pm15$ \\
Ionisation & $\log(\xi/{\rm erg\,cm\,s^{-1}})$ & $4.0_{-0.1}^{+0.4}$ \\
Iron abundance & $A_{\rm Fe}$ & $3.4_{-0.7}^{+0.8}$ \\
Reflection fraction & $R_{\rm Ref}$ & $>3.5$ \\
\hline
\end{tabular}
\end{table}

Before considering the variability in detail, we fit the spectra to determine which components are present to inform our models of the likely sources of variability.
We split each of the four stages of the observation into high and low flux, splitting at the mean flux of that stage.
We use 50\,s bins to do this since this is long enough that the lags between bands do not move flux into different sections but short enough that we are still able to resolve the variability.
We perform fits in \textsc{isis} \citep{houck00} version 1.6.2-41 across the full \nustar\ band ($3-78$\,keV) and always include a cross-calibration constant between FPMA and FPMB \citep{madsen15}.
We group the spectra to a minimum signal-to-noise ratio (SNR) of 25, such that small residuals are readily visible, and use $\chi^2$ statistics.

We model the spectrum initially with a disc blackbody (\textsc{diskbb}) plus powerlaw model. This leaves significant residuals around 8\,keV ($\chi^2/({\rm d.o.f.})=1657/355=4.67$ for the low flux state of the third stage/\nustar\ orbit), so we add an additional blackbody, as has been found in other observations of \maxi\ and interpreted as emission from the plunge region \citep{fabian20}.
This reduces the residuals to below 10\% ($\chi^2/({\rm d.o.f.})=520/353=1.47$ for the low flux state of the third stage/\nustar\ orbit) but bumps are still present at around 7 and 20\,keV. We account for these with the addition of a relativistic reflection component (\textsc{relxilllp}, \citealt{dauser10,garcia14}).
We tie parameters for the reflection component between spectra and fit jointly, with the parameters of the disc, blackbody and powerlaw independent between stages and flux states.
This gives fits with no coherent structure to the residuals, which have significant contributions from the remaining calibration differences between FPMA/B, and $\chi^2/({\rm d.o.f.})=3715/3098=1.20$.
Examples of the residuals to each of these models are shown in Figure~\ref{fig:res}; the best-fit parameters are given in Table~\ref{tab:spec} and shown in Figure~\ref{fig:pars}.

The most significant change between high and low flux states is in the normalisation of the powerlaw. The remaining parameters are consistent with being equal between high and low flux, although constraints are often weak.
There are also changes across the full observation. The disc cools; the extra blackbody also decreases in normalisation and possibly temperature; and the powerlaw displays a softer when brighter trend, as is typical for X-ray coronae.
This is compatible with the following findings from the variability spectra that the flaring is largely due to changes in the powerlaw emission.

\subsection{Energy dependence of the variability}

To determine the source of the variability, we consider how the variability changes with energy. We divide the \nustar\ data into several energy bands, chosen to have approximately equal counts.
We use 18 bands, sufficient that the highest energy band is dominated by the power law component of the spectrum.
We divide the \nicer\ data into bands of at least 0.1\,keV width and to satisfy a minimum counts threshold chosen to give similar error bars to the \nustar\ data.

\subsubsection{Variability spectra}

\begin{figure}
\includegraphics[width=\columnwidth]{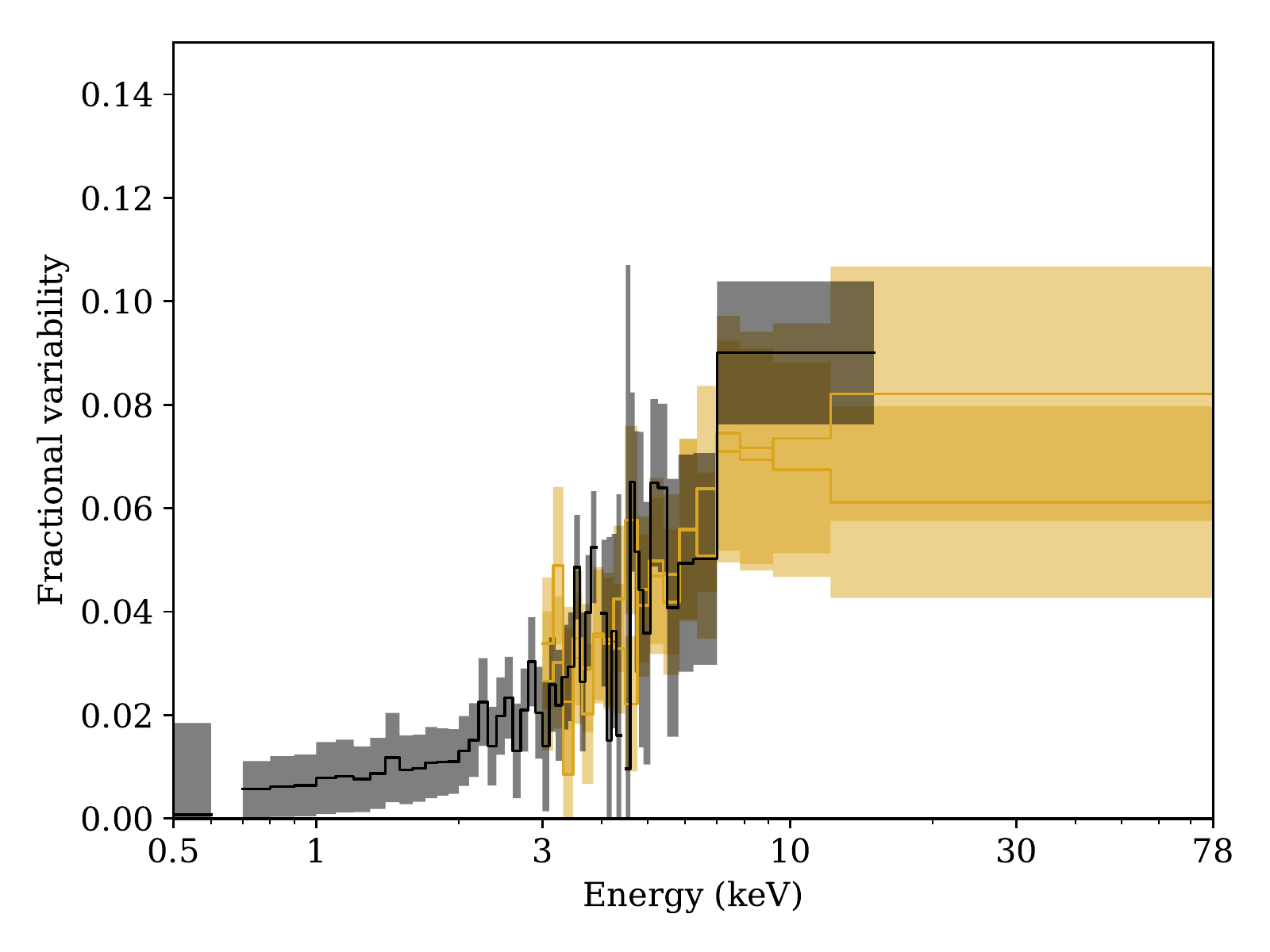}
\includegraphics[width=\columnwidth]{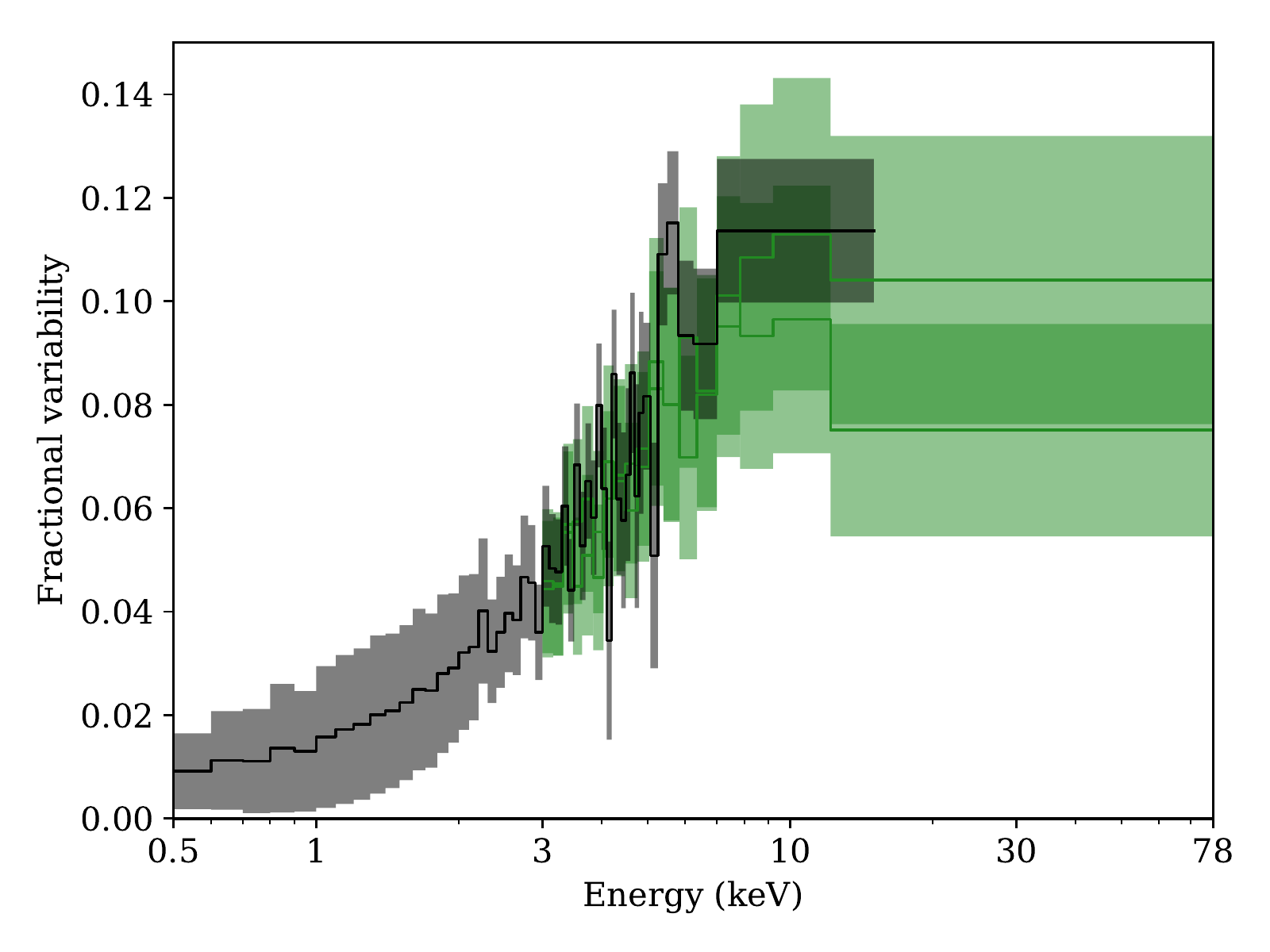}
\caption{Fractional variability spectrum of the variable stages of the observation. Black: \nicer; yellow/green: \nustar. The variability is greatest at high energy, where the spectrum is dominated by the powerlaw component. The flaring (green, lower) phase shows a higher amplitude than but similar shape to the dipping (yellow, upper) phase.}\label{fig:fvar}
\end{figure}

First, we consider how the amount of variability changes with energy.
We measure the fractional variability, after removing the contribution for Poisson noise, using the formulae from \citet{nandra97,edelson02,vaughan03}, for each energy band.
This shows (Figure~\ref{fig:fvar}) that the variability is greatest at high energy, where the emission is dominated by the power-law component. Similarly shaped variability spectra are also found in later observations of the soft state \citep{fabian20}. The spectral shapes of the two stages are very similar but the flaring stage shows a higher amplitude ($\approx9$\% rather than $\approx7$\%).

\subsubsection{Lags}

\begin{figure}
\includegraphics[width=\columnwidth]{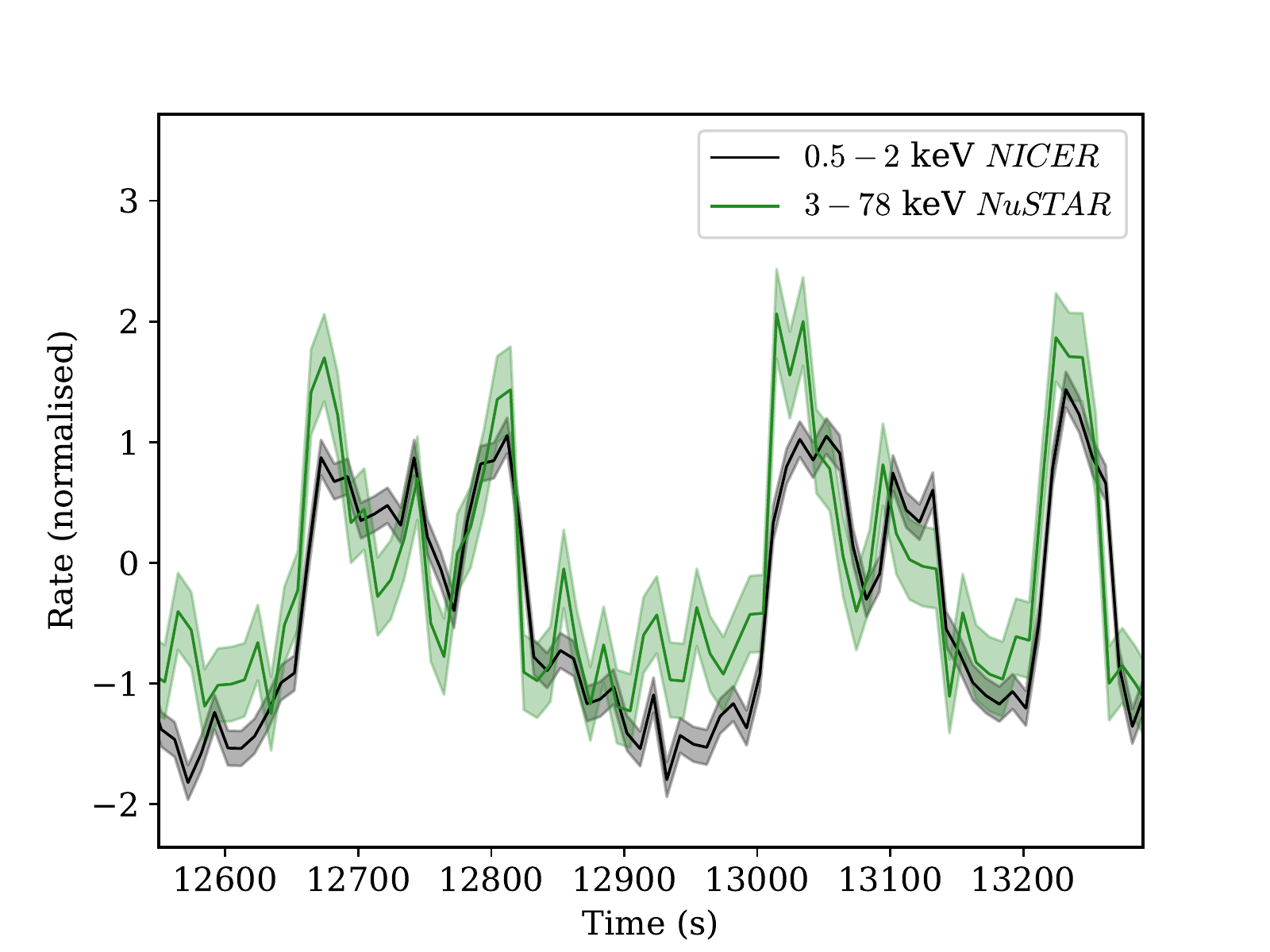}
\caption{Lightcurves in hard (green, $3-78$\,keV \nustar) and soft (black, $0.5-2$\,keV \nicer) bands, zoomed in to a few flares to show that the soft emission lags the hard.}
\label{fig:lc_lags}
\end{figure}

\begin{figure}
\includegraphics[width=\columnwidth]{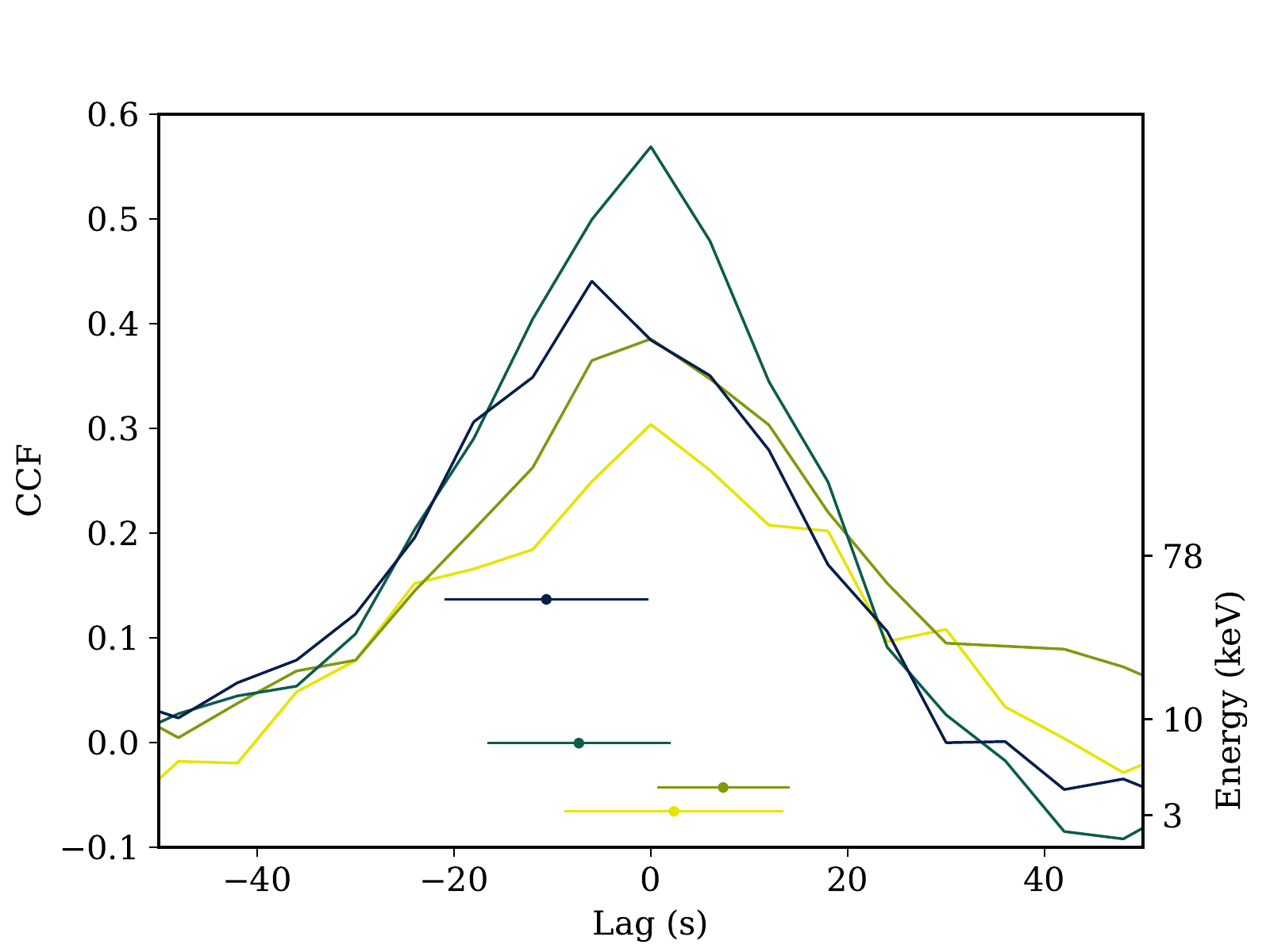}
\caption{Cross-correlation functions (lines, left axis) and centroids (points with errorbars, band energies on right axis) of given energy band against all others, for \nustar\ data.
For clarity, only 4 bands are shown, covering the full energy range; these are the $1^{\rm st}$, $4^{\rm th}$, $9^{\rm th}$ and $12^{\rm th}$ bands in Figure~\ref{fig:lags} (the two lower energy bands have been rebinned by a factor 2 from the resolution used for fitting).
Colours match between CCFs and their centroids: yellow (palest) is softest; blue (darkest) is hardest.
The peak and centroids shift to higher lag in softer bands.
}
\label{fig:ccf}
\end{figure}

\begin{figure}
\includegraphics[width=\columnwidth]{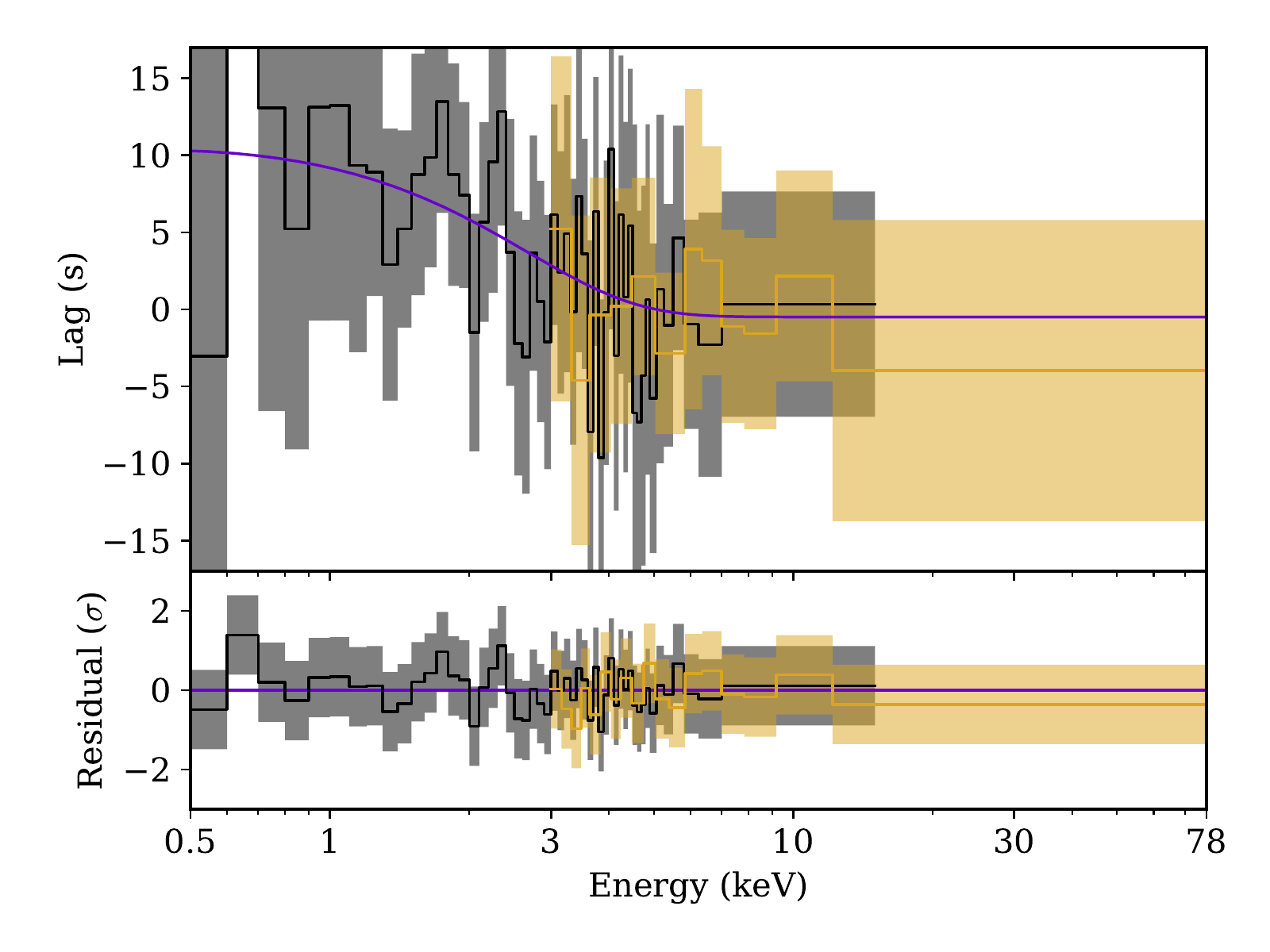}
\includegraphics[width=\columnwidth]{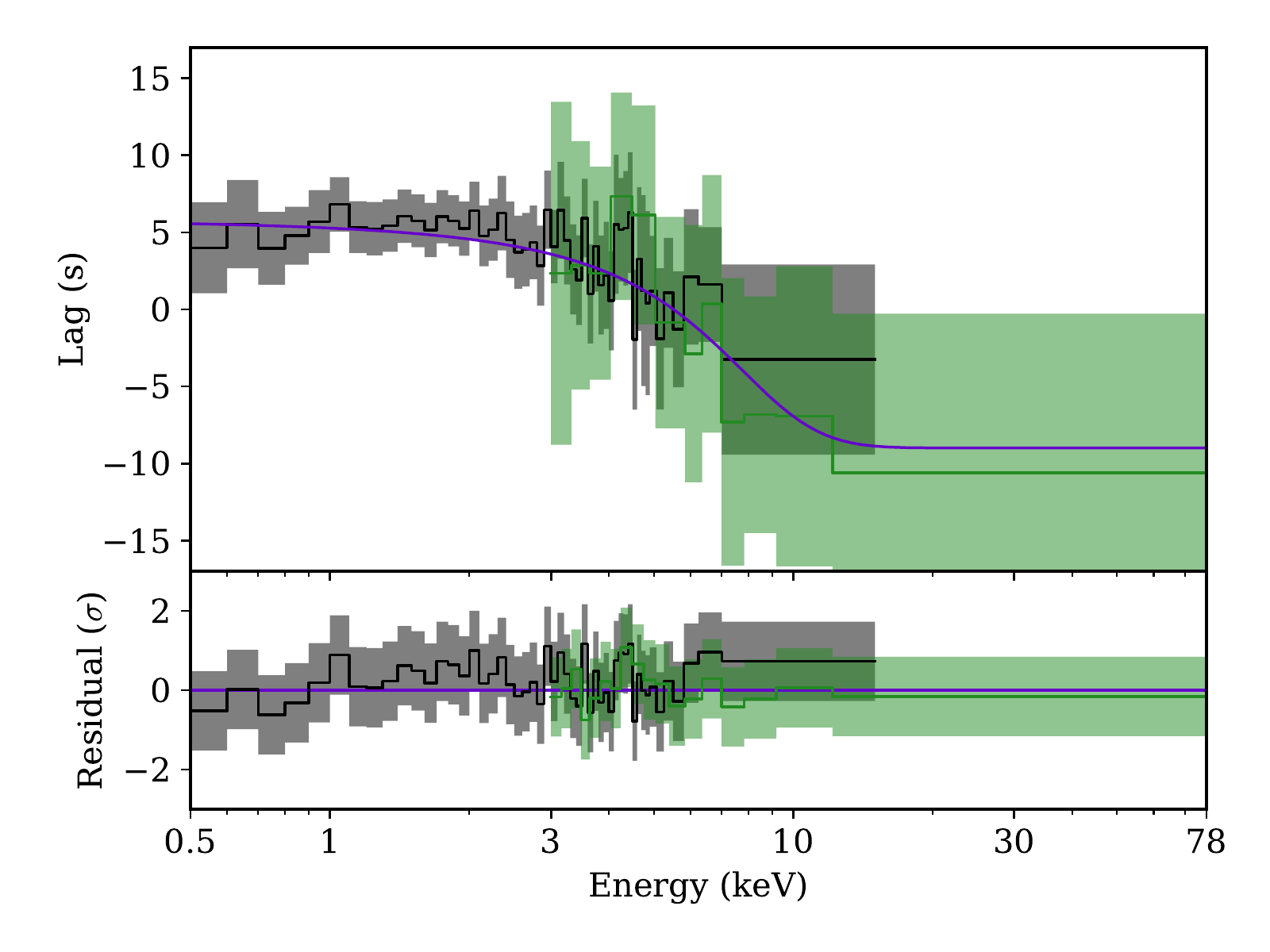}
\caption{Lag-energy plots of the variable stages of the observation.
The lower (green: \nustar, black: \nicer) plot shows the time of flaring variability; this shows a soft lag consistent with being due to a change in the thermal disc component following a change in the coronal powerlaw component. The upper (yellow: \nustar, black: \nicer) plot shows the time of less structured/dipping variability; this shows the lag occurs at lower energies. The first 6 plotted bins of the \nustar\ data have been rebinned by a factor of 2 for clarity.}
\label{fig:lags}
\end{figure}

We also consider the time difference between changes at different energies.
We first show the light curves of soft and hard bands during the structured flaring overlain (Figure~\ref{fig:lc_lags}); this shows that the soft band lags the hard by a few seconds, at least during this stage.
To determine this lag numerically, we use the Cross-Correlation Function (CCF, Figure~\ref{fig:ccf}) because there are relatively few cycles, so Fourier methods are hard to apply reliably.
We calculate the CCF between each band and the rest of the energy range covered by the instrument.
The peak correlation is significantly less than 1 (which would imply that the light curves are linearly correlated at the lag of the peak) and becomes lower at low energies but this does not take into account the dilution of the intrinsic correlation by Poisson noise.
The large number of flares in the light curve means that a spurious correlation is unlikely to arise by chance alignment of independent flares at different energies and, since the physical components emit over a wider energy range than the bands used in the analysis, some correlation between energies is expected.
From the CCFs, we calculate the lag as the centroid of the CCF over the lag range $[-60,60]$\,s. We find errors on these lags using the flux randomisation/subset selection method \citep{peterson98,peterson04}.

The resulting lag-energy spectrum is shown in Figure~\ref{fig:lags}. This shows a soft lag (i.e. changes at lower energies happen later) in the bands covering $\sim3-78$\,keV, which flattens off at lower energies.
This corresponds to a flat lag-energy profile where the spectrum is disc-dominated and progressively earlier changes in emission as the Comptonised component becomes more significant.
To demonstrate this numerically, we fit this with a simple model including only changes in the normalisation of a thermal and a power law component (we consider more detailed models in Section~\ref{sec:specflare}).
We approximate the spectrum of each variable component (thermal or power law) as the equivalent component in the mean spectrum multiplied by its fractional variability, effectively requiring that the coherence is independent of energy.
Due to the lower signal-to-noise than the mean spectrum we do not include both the disc and secondary black body; rather, we use a single thermal component with power equal to the total of the two components used in the mean spectrum.
Further, we allow the temperature, $kT$, of this component to be a free parameter, such that the energy at which the sloped part of the lag-energy relation occurs can be fitted for.
The lag is introduced by allowing the change in thermal emission (with flux change $df_{\rm Thermal}(E,kT)$) to occur a time $\Delta t$ after the change in power law emission ($df_{\rm PL}(E)$).
The mean time of arrival (and so the measured lag) in a given band is then 
$$t(E) = t_0 + \Delta t \frac{df_{\rm Thermal}(E,kT)}{df_{\rm Thermal}(E,kT)+df_{\rm PL}(E)}$$
where $t_0$ is the zeropoint.
We then fit for $t_0$, $\Delta t$ and $kT$. We use $\chi^2$ statistics.
This fits the overall shape well.
For the flaring stage, the temperature of the thermal component, $1.11\pm0.03$\,keV, is between the temperatures of the disc and secondary thermal component found from spectral fitting (see Section~\ref{sec:spec}).
The time difference of $20.0_{-1.2}^{+1.6}$\,s corresponds to around 10\% of the typical flare recurrence time (which is the dominant variability time scale).
For the dipping stage, the temperature, $0.53^{+0.03}_{-0.08}$\,keV, is significantly lower than that observed in the flaring stage: forcing the temperature to match that of the flaring stage produces $\Delta\chi^2=15$. It is also lower than any of the thermal components found in the spectral model. Thus, this model predicts that the lag occurs some way out in the disc. A more likely scenario could be that the process which produces the lag acts across all radii and becomes longer at larger radii (lower energies). The longer lag at lower energies produces a sloped lag-energy profile which is similar to a lower disc temperature. Due to the variety of possible lag-radius relations and limited signal, we do not attempt to fit a more complex model.

Therefore, the lag-energy spectra imply that the lag is caused by a change in the disc which occurs after the change in the Comptonised emission. For example, two possible changes are in the inner emitting radius ($R_{\rm in}$) or the local disc temperature at a given radius ($T|_R$). In the following section, we test what type of change is supported by the data,
before proposing a likely physical scenario in Section~\ref{sec:phys}.

\subsection{Spectral analysis -- flares}
\label{sec:specflare}

\begin{figure*}
\includegraphics[width=\columnwidth]{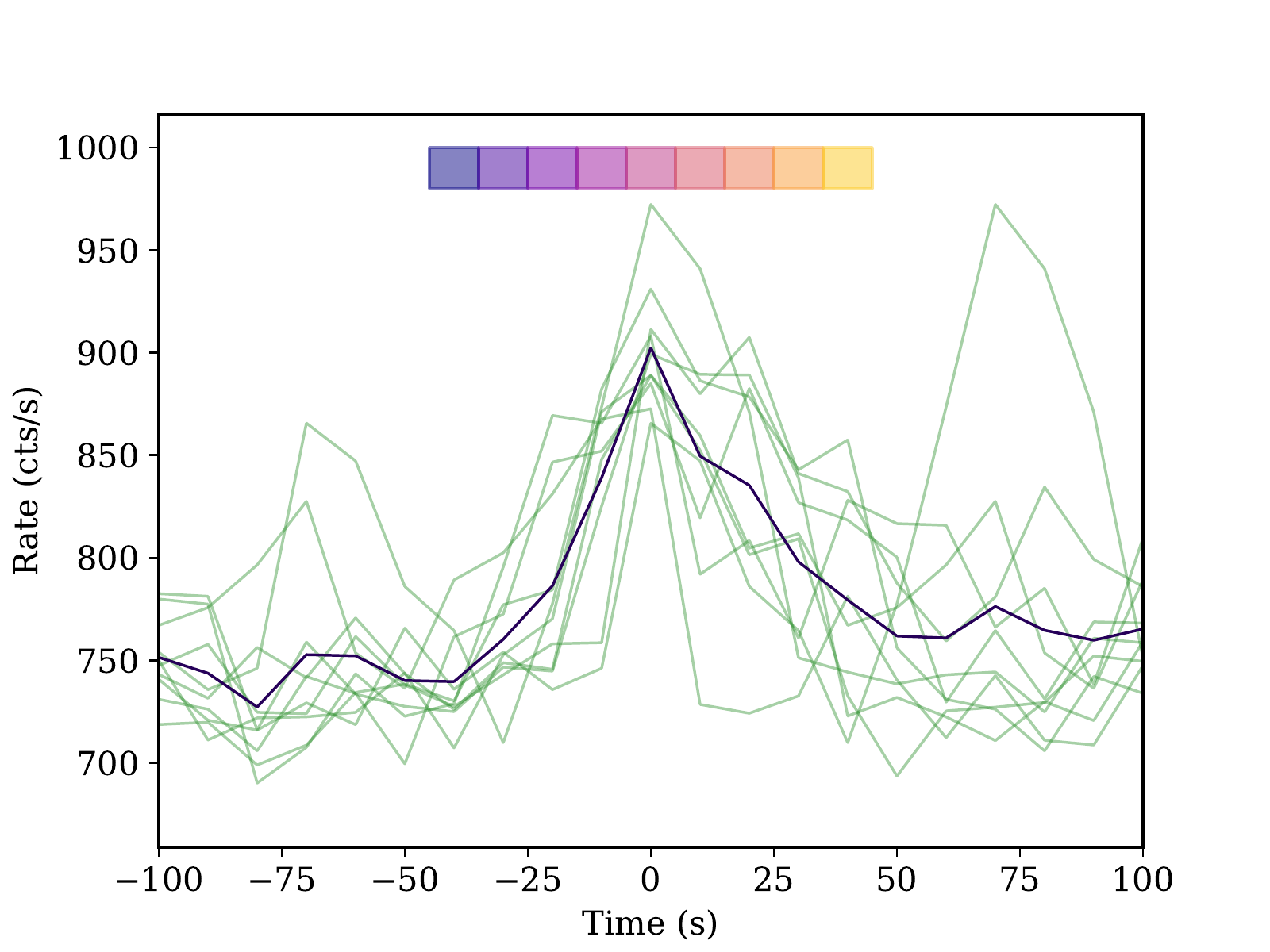}
\includegraphics[width=\columnwidth]{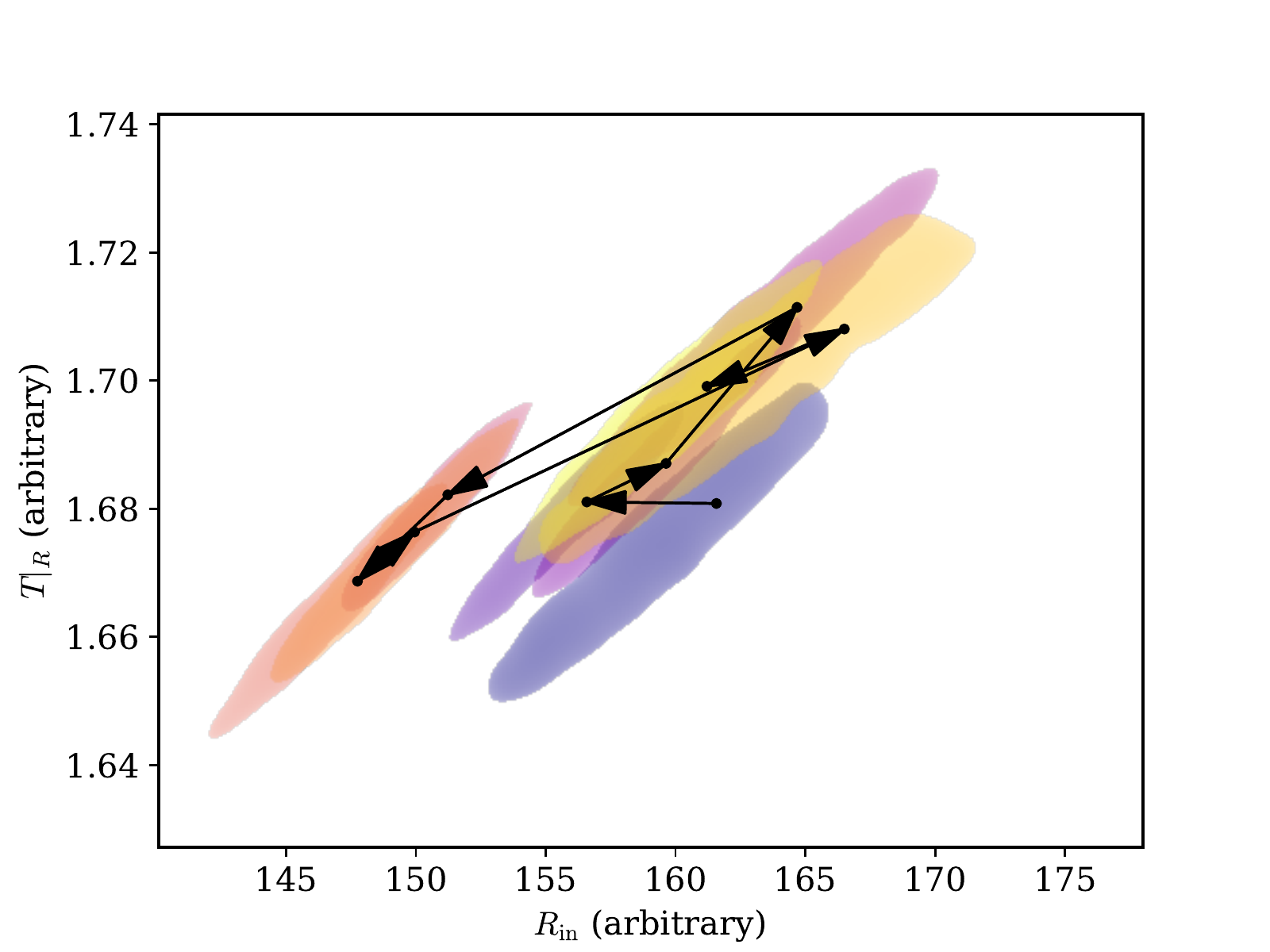}

\includegraphics[width=\columnwidth]{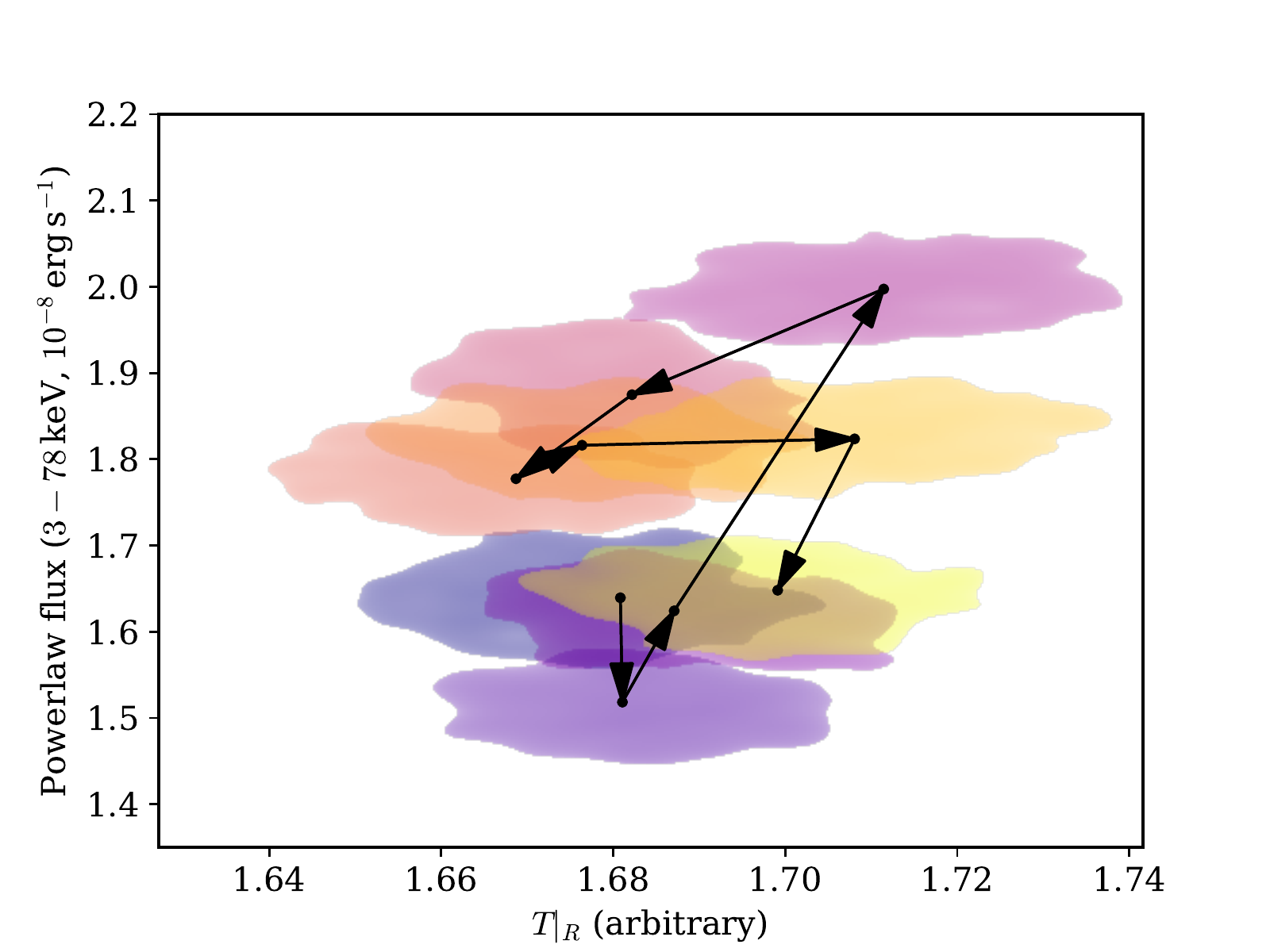}
\includegraphics[width=\columnwidth]{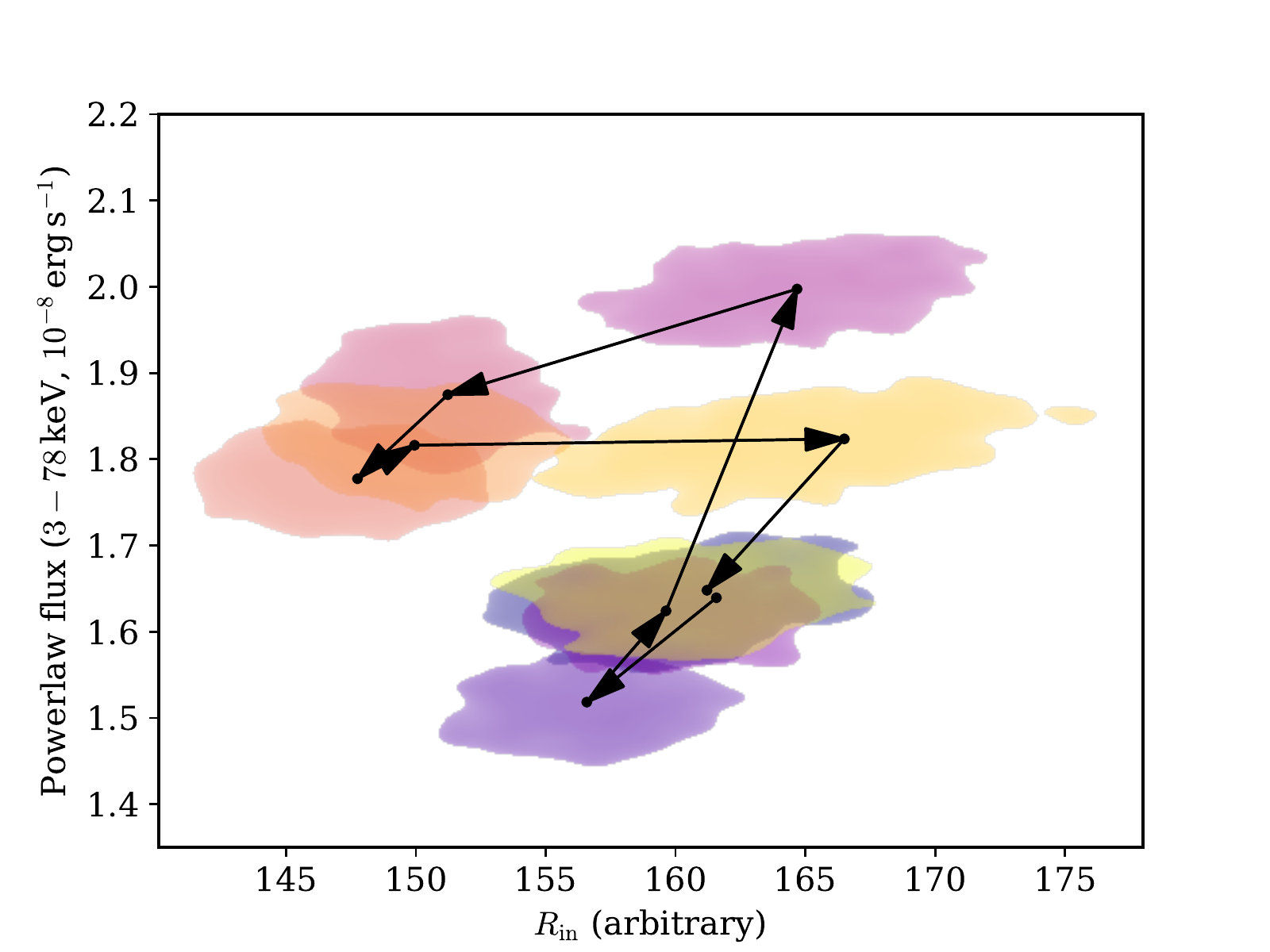}
\caption{
\textit{Top left:} Each flare shifted to peak at $t=0$ (green) and their mean (purple). The coloured patches at the top indicate the times used to extract spectra and match the colours in the remaining panels.
{\it Remaining panels:} parameter evolution of fits to stacked flare spectrum. Arrows indicate the direction of time (the highest count rate occurs one bin after the power law peak and one before the $R_{\rm in}$ minimum).
\textit{Top right:} the disc shows a change in the  $R_{\rm in}$-$T|_R$ plane with a potential change in $R_{\rm in}$ and at most a small change in temperature at a given radius, $T|_R$ (shaded regions include 68\% of the posterior density).
\textit{bottom left:} any change in $T|_R$ is small, but consistent with rapid re-radiation of illumination from the corona.
\textit{bottom right:} the smallest $R_{\rm in}$ (and hence strongest disc emission) occurs after the peak in powerlaw flux, so is the best candidate to be responsible for the soft lag.
}
\label{fig:peakstack}
\label{fig:peakpars}
\end{figure*}

We may also try to find the changes directly in time-resolved spectra. There is insufficient signal to place useful constraints across any individual change in flux, so some stacking is required.
Since the variability in the third \nustar\ orbit is dominated by flaring events, we stack these flares (the variability during the dipping stage is less structured, so we cannot easily perform a similar analysis for that stage), with the caveat that the flares differ in light curve (Figure~\ref{fig:peakstack}), so the stacking adds slightly different spectra in each time bin.
We identify flare peaks as points in the light curve which are above 850\,cts/s and higher than their nearest two neighbours before and after (i.e. 4 nearest neighbours in total). These cuts are chosen to avoid smaller fluctuations which appear to be part of a larger flare.
We align the flares at their peaks and extract spectra in time segments of 10\,s, summing each segment over all flares.
Since \nustar\ observes the high energy tail of the disc, there is a significant degeneracy between the inner radius and temperature. To reduce this, we also include the \nicer\ data. Owing to the soft spectrum and high soft throughput of \nicer, this can lead to the powerlaw component fitting to small deviations from the disc model rather than the true high energy Comptonised emission.
Therefore, we include a 2\% systematic error on the \nicer\ data, sufficient that the powerlaw fits the high energy data well.
We fit the resulting spectra with the model from Section \ref{sec:spec}. Due to the limited SNR available, we freeze the reflection parameters to the best fit values from Section \ref{sec:spec}. This is not ideal, since during this highly variable interval these parameters may change. The focus on changes in the disc parameters is justified by the inference from the lag-energy spectra that the lags are caused by delays between disc and coronal changes. Further, the disc temperature structure may not match the equilibrium profile used in \textsc{diskbb}.
We use a Markov Chain Monte-Carlo (MCMC) run (with 175 walkers and 430 steps after a burn-in period of 570 steps) such that we may directly explore parameter degeneracies. We use the implementation in ISIS of the implementation of \citet{foremanmackey13} of the method of \citet{goodman10}.
We use flat priors on each parameter, with ranges based on the fits from Section \ref{sec:spec}.

We show the changes in important pairs of parameters in Figure~\ref{fig:peakpars}; the coloured regions show the 68\% of the posterior weight with highest density.
As expected, the most prominent change is the increase in powerlaw flux during the flare. We also expect a change in the disc parameters reflecting a peak in disc emission after the peak in powerlaw emission. This could be due to a change in the inner radius ($R_{\rm in}$) or in the temperature at a given radius ($T|_R$, reflecting the combination of mass accretion rate and re-radiation of illuminating power).
Several points after the powerlaw peak are clearly offset in the $R_{\rm in}$-$T|_{R}$ plane. However, due to the degeneracy between $R_{\rm in}$ and $T|_{R}$, identifying this as due to one of these parameters is less clear.
We also consider whether the changes relative to the powerlaw agree with the disc change lagging.
While any change in $T|_R$ is poorly constrained, it is consistent with being linearly dependent on powerlaw flux, as would be expected from re-radiation of the coronal illumination of the disc (which should occur much more rapidly than the timescales probed here).
Contrastingly, the apparent decrease in $R_{\rm in}$ occurs after the peak in powerlaw emission, forming an anticlockwise loop in the $R_{\rm in}$-$T|_{R}$ plane. Since flux is greater at smaller $R_{\rm in}$, this hysteresis  agrees with the observed lag.

Therefore, a change in the inner radius being responsible for the lag is supported by the data, although the simplifications and limited SNR mean that this is not unambiguously proved from the data alone.

\subsection{Physical interpretation and checks}
\label{sec:phys}

Based on the lag-energy spectra and the suggestions from the time resolved spectral fitting, we propose the following scenario for the observed variability properties.

During the dipping phase, we observe that the powerlaw and inner disc change together and lower energies change later.
This could be initiated by a drop in inner disc density and emission (which is also powering the coronal emission). This is followed by a rarefaction wave passing outwards through the disc, causing a drop in flux at progressively larger radii and lower energies, so generating the observed lags.
During the flaring phase, there may be a more extreme disruption of the inner disc, ejecting material/power into the corona and causing a small increase in the inner disc radius; the disc then refills, restoring the disc emission and producing the observed lags.
It is not clear what has changed in the system that such different processes occur but the phenomenological difference in the light curve clearly shows that there is a difference between the two variability cases.

We also test that these proposed changes are consistent with the timescales expected for accretion discs.
In this section, we use a mass $M=9.2\,{\rm M_{\odot}}$ (the middle of the range from \citealt{atri20}) and an Eddington ratio $\dot{m}=0.2$ (using the luminosity of $2.26\times10^{38}$\,erg\,s$^{-1}$ from \citealt{fabian20}). To use the calculations of \citet{shakura73}, we use an inner radius $R_{\rm in}=6r_{\rm g}$, matching the ISCO of a zero spin black hole; this is reasonable since \citet{buisson19} find a low spin.

We first consider the proposed disruption/refilling during the flaring (green) variability.
From equations~2.11,16,19 of \citet{shakura73}, we integrate the radial (inward) velocity from $1.1R_{\rm in}$ to $R_{\rm in}$ (the fractional change in radius found in Figure~\ref{fig:peakpars} and assuming that the minimum observed $R_{\rm in}$ is $r_{\rm ISCO}$) and hence the time for the disc to refill by accretion. This time is inversely proportional to $\alpha$, the viscosity parameter; we find that $\alpha\approx6\times10^{-4}$ gives a refilling time of 20\,s, matching the observations.
This $\alpha$ is rather smaller than the common fiducial value $\alpha\approx0.1$ \citep{king07,martin19}, although most previous empirical constraints giving $\alpha\gtrsim0.1$  (such as modelling long-term outburst light curves, e.g. \citealt{dubus01,kotko12,tetarenko18}) are dominated by the outer regions of the disc, where the viscosity may differ from the innermost regions probed here. 
Optical variability in AGN, which is generated at intermediate disc radii, can imply a slightly lower $0.01\leq\alpha\leq0.03$ \citep{starling04}, which supports the idea that $\alpha$ is smaller at smaller radii.
Also, our value does agree with some early estimates of the viscosity: \citet{shakura73} suggest $10^{-3}$ may be suitable close to the inner radius; and arguments on the Reynolds number of the flow by \citet{lyndenbell74} imply $\alpha\lesssim10^{-3}$.
More recent simulations of discs also tend to give values $\alpha\lesssim0.02$ (e.g. \citealt{stone96,hirose06} but see \citealt{hawley01}).

We also propose that a wave travelling outwards provides the lags during the dipping (yellow) variability.
We find a similar lag time ($\mathcal{O}(10)$\,s) over a factor $\mathcal{O}(10)$ in energy. Since $E\propto T\propto r^{-3/4}$, the change in radius associated with this is also $\mathcal{O}(10)$, or $\mathcal{O}(100)$ times larger than the radius range involved in the flaring variability. Such a wave is likely to travel at around the sound speed, governed by the thermal timescale, as opposed to the viscous timescale for disc refilling.
The thermal timescale is rather faster, $t_{\rm th}\sim (h/R)^2 t_{\rm v}$ \citep[e.g.][]{pringle81}, so the required $\mathcal{O}(100)$ speed increase would be associated with a plausible $(h/R)\approx0.1$ \citep[e.g.][]{mcclintock06}.
Overall, within the current understanding of accretion disc theory, the required timescales for the processes we propose are not impossible.

\section{Discussion}
\label{section_discussion}

We have described a brief period of strong variability late during the hard to soft state transition of the X-ray binary \maxi.
The variability occurs principally in the flux of the Comptonised emission, as is typically seen in XRBs \citep[e.g.][]{belloni05}, but unusually shows a long soft lag, of $20.0_{-1.2}^{+1.6}$\,s.
The state transition is, as with most XRBs, short compared with the full outburst, so datasets from pointed, sensitive telescopes covering transitions are comparatively rare and valuable; therefore, strong, transient variability such as that observed here may be more common than is apparent from the existing literature.

The flaring we observe has more regular structure in the light curve than the broadband noise variability which is most often observed in XRBs.
Another qualitatively similar phenomenon is the Type~II X-ray bursts observed in some accreting neutron stars \citep{hoffman78,fishman95,kouveliotou96}.
The occurrence of Type~II bursts is rather different to the flares observed here: they are observed in very few sources (2 of over 100 known neutron star low mass XRBs) but frequently in the sources where they are seen \citep[e.g.][]{court18}.
Any flaring behaviour in an X-ray binary can be compared with V404~Cyg \citep[e.g.][]{gandhi16,walton17,motta17abs}, though again the flares in V404~Cyg are different to those observed here: they have a wide variety of shapes, strengths and recurrence times, whereas the flares in \maxi\ are very regular.

\label{sec:ff}

\begin{figure}
\includegraphics[width=\columnwidth]{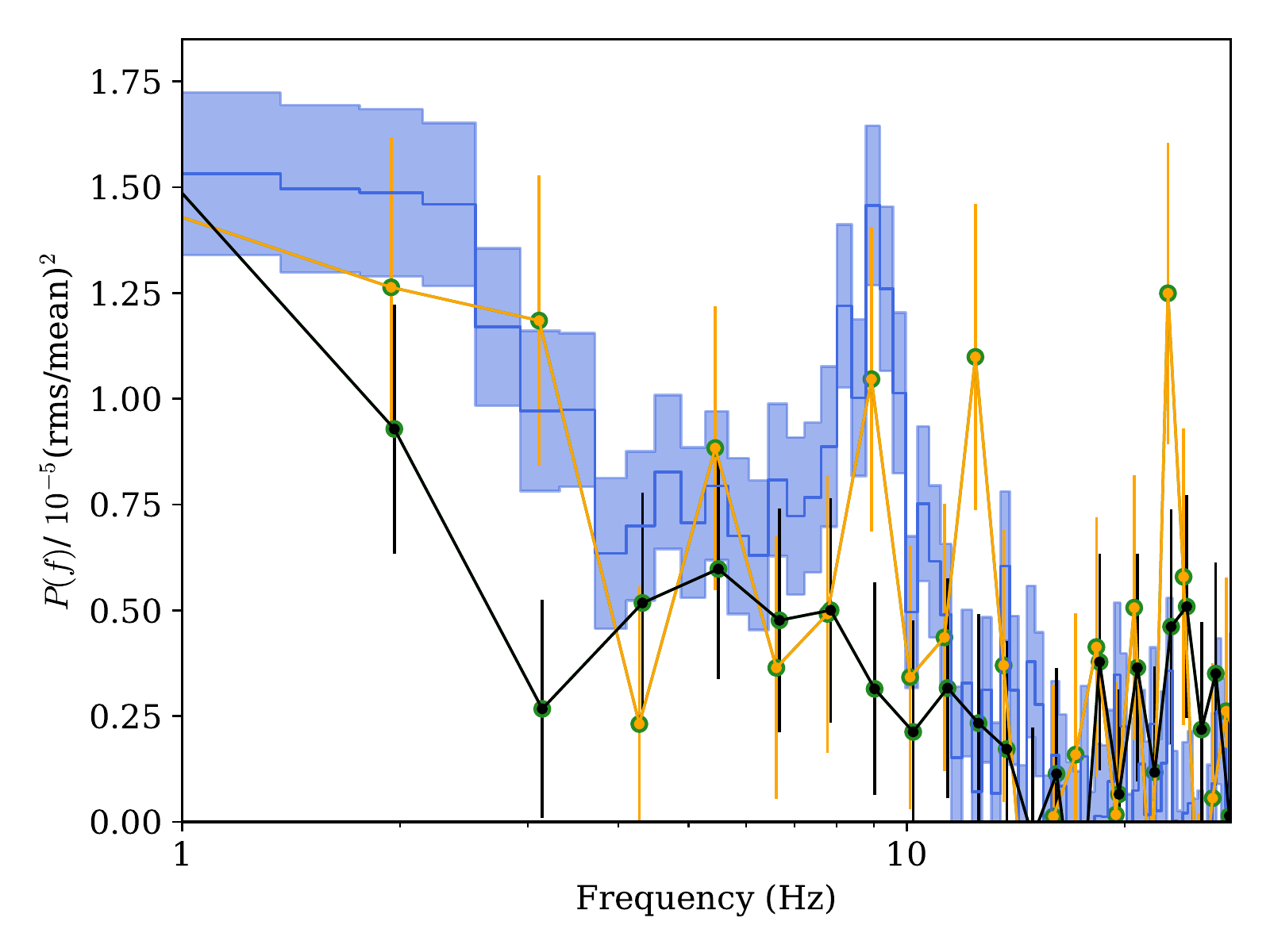}
\caption{The 9\,Hz QPO found in the plateau following the strongly variable section (blue), which is consistent with the power spectrum of the low phase during the flaring (orange/green) but not the high phase during the flaring stage (black/green). The power spectra shown are calculated from 0.5-10\,keV \nicer\ data.}
\label{fig:qpo}
\end{figure}

A more similar type of behaviour is that of `flip-flop' phases \citep{miyamoto91}, which have been seen in a small number of sources.
These are characterised by repeated transitions between the same two states; the transition between the states is typically much faster than the duration in each state.
from a few tens of seconds \citep{miyamoto91} to several thousand seconds \citep{bogensberger20}. Slower changes may exist but this makes observing sufficient changes to recognise these as a `flip-flop' less likely \citep[e.g. compare][]{xu19,bogensberger20}.
They may also involve changes between QPO types in the different states \citep[e.g.][]{casella04}, although the QPO types involved are not consistent between cases. 
There are also different dominant spectral changes; for example a significant change in disc temperature may \citep[e.g.][]{bogensberger20} or may not \citep[e.g.][]{miyamoto91} be observed. 
All previous cases have occurred while the source was in an intermediate state, i.e. when there was significant contribution from both disc and coronal emission.
Also, a QPO is usually seen in at least one of the states.
There is a QPO in the steady state following the flaring (Figure~\ref{fig:qpo}, this is also mentioned briefly in \citealt{homan20}); this interval (blue in Figure~\ref{fig:lc}) is very similar in spectrum and flux to the low phase of the flaring. The power spectrum during the low state of the flaring stage is consistent with the QPO occurring then also, although there are not enough data to confirm this; the QPO is not present during high flux phases of the flaring stage (or the earlier steady stage).
Therefore, though the full range of this `flip-flop' behaviour is still poorly defined, the flaring in \maxi\ agrees with the spectral and temporal properties of this class.
\citet{phillipson18} note the similarity of this behaviour to the Duffing oscillator, a double welled (hence two typical flux states) non-linear (hence chaotic) oscillator. However, they do not determine the properties of the binary system which correspond to the variables in the Duffing equation:
there is not yet a physical explanation for `flip-flop' behaviour, so this identification still leaves the physical change to be determined.

Various forms of highly structured variability are also seen in GRS~1915+105.
A structure particularly similar to the light curve shown here was analysed by \citet{belloni97}, which showed a high steady phase separated from a low steady phase by a period of rapid variability. Several such cycles were observed in GRS~1915+105; while only one is observed in \maxi\ it is not impossible that others occurred outside times of observation.
\citet{belloni97} suggest that this behaviour is due to the Lightman-Eardley instability \citep{lightman74}, although since this instability is always expected to occur during bright phases (where the disc includes a region with higher radiation than gas pressure) of X-ray binary outbursts, it is unclear why this variability pattern is only rarely seen. One possibility is that a fine-tuned balance with another stabilising mechanism is required.

A further source of information is the time lag between changes at different energies: the lags we find are unusual in both their sense and duration.
Most lags observed in X-ray binaries have the hard emission lagging the soft \citep[e.g.][]{uttley11} and soft lags are seen at high frequencies with correspondingly short magnitude \citep[typically milliseconds, e.g.][]{kara19}.
Conventional soft lags are explained by reflection of coronal emission from the disc which lags the direct coronal emission by the light travel time from the corona to disc.
However, the light travel distance associated with the lag observed here is $\sim6\times10^{11}$\,cm, many orders of magnitude larger than the X-ray emitting disc. Reprocessing from X-ray to optical can produce lags of the required magnitude \citep[e.g.][]{gandhi10} but these have been found to be slightly shorter in \maxi\ \citep[$\approx5$\,s][]{paice19} due to the smaller disc; hard X-ray to soft X-ray reprocessing is expected to give even shorter lags. Therefore, reprocessing in the disc is unlikely to be the primary source of lags.
The light travel distance is a good match for the distance to the companion but the scattered emission from the companion star does not have the correct spectrum (which the lag-energy spectrum shows can be approximated by a $\sim1$\,keV thermal component) and would be too weak to produce the observed lags.
A change in the underlying accretion system seems a more likely explanation.

The variability in \maxi\ analysed here occurs close to but distinct in time to several other features in the state change. There may be some association with the jet: around 1 day before the flaring variability, there is a jet ejection coincident with a type B QPO \citep{bright20,espinasse20,homan20}.
The flaring is too late to be directly associated with the ejection but could represent settling from the post-ejection state to an equilibrium.
We note that a strong increase in radio flux was also seen before the onset of the `flip-flops' in Swift~J1658.2--4242.
The flaring may be a small-scale manifestation of the jet cycle \citep[e.g.][]{lohfink13} possibly due to settling after the ejection event. 
This agrees with the cooling and increase of inner radius at the start of the flare, as though the inner disc was ejected, and the later decrease in inner radius and increase in temperature, which could be due to the inner disc refilling.
The flaring variability occurs during a gap in the radio coverage presented in \citet{homan20} but the radio flux around 0.5 days later is low, so this cannot form a new strong jet, although the flares are much smaller than the original state change so this would not necessarily be expected.
A piece of evidence against this is that similar flaring was also observed in an outburst of 4U~1630$-$47, which showed no significant radio activity \citep{tomsick05}.
This could be because the bimodal flaring is not directly related to the radio activity, but requires a specific narrow set of parameters which can be induced after a jet ejection; perhaps in 4U~1630$-$37, the relatively complex outburst profile allowed the required conditions to be met.

Long duration soft lags have also been observed in ULXs \citep{heil10, demarco13ulx,hernandez15,pinto17,kara20ulx}, where they appear to be associated with the hard component of emission; like \maxi, they typically show a lag of around 10\% of the variability timescale probed.
There is no consensus on the reason for these lags.
If the lags observed here are manifestations of the same phenomenon, this would imply that there is a similar causal connection between the hotter and cooler thermal components seen in ULX spectra as is present between the more standard disc and corona in MAXI J1820.

\section{Conclusions}
\label{section_conclusions}

\begin{itemize}

\item We find a period of strong variability between two plateaus in flux in the latter stages of the hard to soft state transition of \maxi. As is typical for XRBs, the time for the transition to occur is a small fraction of the total outburst length, so this is a comparatively rare dataset.

\item The first orbit of the variable epoch shows dipping with relatively little long-term structure.

\item The second orbit consists of flares from the eventual plateau flux to around 25\% brighter.
These flares increase in frequency through the orbit. Unfortunately, the eventual assumption of the final plateau flux occurs during an orbital gap.

\item The variability is primarily due to changes in the flux of the powerlaw emission; there is also a smaller change in the disc emission.

\item We also find a soft lag of the thermal peak behind the powerlaw flux of $20.0_{-1.2}^{+1.6}$\,s.

\item The emission components must also change spectral shape. Fitting time-resolved spectra around the flare peaks suggests that this is a change in the disc temperature and inner radius, consistent with the change being governed by removal/refilling of the inner disc, with the temperature at a given radius remaining constant.

\item We suggest that the lags during the dipping phase are from a sonic wave passing outwards through the disc, while in the flaring phase, the inner disc is disrupted during the coronal flare and then refills viscously. This is consistent with the timescales expected from classical accretion disc theory for a relatively low $\alpha\lesssim10^{-3}$.

\item These oscillations between disc and coronal emission could be due to the disc/corona power balance returning to its ground state after being perturbed by the jet ejection $\sim1$\,day earlier.

\end{itemize}

\section*{Data availability}

The data on which this work is based are available from HEASARC.

\section*{Acknowledgements}

We thank the referee for helpful comments, which improved the manuscript.
We thank Javier Garcia, Adam Ingram, Michiel van der Klis and Guglielmo Mastroserio for helpful discussions.
We thank Fiona Harrison for approval of these DDT observations and Karl Forster for their prompt scheduling.
DJKB acknowledges financial support from the Science and Technology Facilities Council (STFC) and the Royal Society.
DJW acknowledges support from STFC in the form of an Ernest Rutherford fellowship.
This work made use of data from the \nustar\ mission, a project led by the California Institute of Technology, managed by the Jet Propulsion Laboratory, and funded by the National Aeronautics and Space Administration. This research has made use of the \nustar\ Data Analysis Software (NuSTARDAS) jointly developed by the ASI Science Data Center (ASDC, Italy) and the California Institute of Technology (USA).
This research has made use of ISIS functions (ISISscripts) provided by ECAP/Remeis observatory and MIT (http://www.sternwarte.uni-erlangen.de/isis/).

\bibliographystyle{mnras}
\bibliography{maxij1820}

\bsp	% typesetting comment
\label{lastpage}
\end{document}